\documentclass[aps,prl,groupedaddress,twocolumn,superscriptaddress]{revtex4-2}
\usepackage{amsmath,amssymb,mathrsfs,amsfonts}
\usepackage{txfonts}
\usepackage{upgreek}
\usepackage{bm}
\usepackage{ragged2e}
\usepackage{color}
\usepackage{placeins}
\usepackage{float}
\usepackage{appendix}
\usepackage{color}
\usepackage[para]{threeparttable}
\usepackage{epsfig}
\usepackage{threeparttable}
\usepackage{epstopdf}
\usepackage{mfirstuc}
\usepackage[colorlinks,
            citecolor=blue,
            anchorcolor=red,
            menucolor=red,
            linkcolor=blue,
            filecolor=red,
            runcolor=red,
            urlcolor=blue,
            frenchlinks=red]{hyperref}
\makeatletter
\providecommand{\fu}[1]{\textcolor{blue}{#1}}

\newcommand{\Rmnum}[1]{\expandafter\@slowromancap\romannumeral #1@}
\makeatother
\begin{document}

  \title{Ellipticity-Controlled Bright-Dark Coherence Transition in Monolayer WSe$_2$}

\author{Kang Lan}
\affiliation{School of Physics and Physical Engineering, Qufu Normal University, 273165, Qufu, China}
\author{Xiangji Cai}
\affiliation{School of Science, Shandong Jianzhu University, Jinan 250101, China}
\author{Zhongxiao Man}
\affiliation{School of Physics and Physical Engineering, Qufu Normal University, 273165, Qufu, China}
\author{Shijie Xie}
\affiliation{School of Physics, State Key Laboratory of Crystal Materials, Shandong University, Jinan 250100, China}
 \author{Ning Hao}
\affiliation{Anhui Province Key Laboratory of Low-Energy Quantum Materials and Devices , High Magnetic Field Laboratory, HFIPS, Chinese Academy of Sciences, Hefei, Anhui 230031, China}
\author{Ping Zhang}
\affiliation{School of Physics and Physical Engineering, Qufu Normal University, 273165, Qufu, China}
\affiliation{Beijing Computational Science Research Center, Beijing 100084, China}
\author{Jiyong Fu}
\thanks{yongjf@qfnu.edu.cn}
\affiliation{School of Physics and Physical Engineering, Qufu Normal University, 273165, Qufu, China}

\begin{abstract}
  The generation of exciton valley coherence typically requires linearly polarized (LP) light as an external coherent drive, whereas circularly polarized (CP) light fails to induce coherence. 
  Here, we develop a unified, microscopically-grounded open-quantum-system framework within a five-level model incorporating bright-dark exciton interactions in monolayer WSe$_2$, and demonstrate that the polarization ellipticity of the excitation field provides selective control over distinct exciton species contributing to valley coherence. Specifically, LP and CP excitations generate bright and dark coherence, respectively, with continuous \emph{ellipticity tuning} enabling controlled transitions between these states.
  We further reveal dual magnetic advantages for manipulating dark coherence even in the absence of initial coherence: (i) an out-of-plane magnetic field suppresses coherence decay and (ii) an in-plane field enables its optical readout, with quantitatively realistic field strengths.
  These findings provide a powerful mechanism for accessing hidden dark states via ellipticity-driven coherence transfer, and establish a new pathway for harnessing
  bright-dark valley-coherence transitions  in future quantum control.

\end{abstract}

\maketitle

\textit{Introduction}.
Monolayer transition-metal dichalcogenides (TMDCs) $MX_2$ ($M$=Mo, W; $X$=S, Se) host two inequivalent valleys ($K$ and $K^\prime$) connected by time-reversal symmetry, featuring spin-valley locking and valley-selective optical excitation using circularly polarized (CP) light~\cite{PhysRevLett.108.196802,Ye2014,PhysRevB.77.235406,wangkalantar2012,2whm-6xyv,sciadv.adr5562}.
Strong Coulomb interactions in these atomically thin semiconductors lead to tightly bound excitons that serve as an ideal platform for exploring excitonic valley coherence (VC)--a quantum superposition between valley excitons enabling optical initialization and quantum-state control~\cite{PhysRevLett.120.046402,ElHouri2024,9spx-3hrq,PhysRevB.108.035419,selig2016excitonic,PhysRevX.14.011004,nanolett.2c04536,nanolett.4c01327,PhysRevB.104.L121408}.

Existing VC studies mainly rely on bright excitons driven by linearly polarized (LP) light, yet incoherent intervalley scattering and rapid radiative  decay pose fundamental challenges for coherent valley manipulation~\cite{jones_yu2013,PhysRevLett.117.187401,qiu2019room,hao_moody2016,moody2015intrinsic,gupta2023,PhysRevB.110.125420,acsnano5c02659}.
In contrast, dark excitons have recently emerged as promising candidates for valley-based qubits and coherence control in TMDCs, owing to their longer lifetimes and \emph{robust} intervalley exchange interaction~\cite{PhysRevLett.123.096803,PhysRevB.96.155423,PhysRevLett.131.116901,scienceaba1029}.
Previous works have established their existence, optical accessibility, and coherence properties primarily via magnetic-field-induced brightening~\cite{PhysRevLett.123.096803} and time-resolved spectroscopy~\cite{PhysRevB.96.155423}, focusing on coherence within the dark manifold once the states are prepared.
However, the mechanisms for optically generating and manipulating dark coherence remain elusive, particularly given that dark excitons typically arise indirectly via intravalley scattering from bright states.
Moreover, how external magnetic fields can further enhance or enable optical readout of dark coherence remains an open and compelling question.

In this letter,  demonstrate how polarized light--CP, LP, and elliptically polarized (EP)--selectively governs which exciton species (bright or dark) participate in valley coherence through bright-dark exciton coupling.
Specifically, we reveal that dark coherence spontaneously emerges under CP excitation through exchange-mediated intervalley repopulation,  activated by valley population imbalance, whereas bright coherence requires an initial coherent drive provided by LP light.
This selectivity originates from ellipticity-dependent \emph{initial} state preparation, allowing exciton coherence generation even without preexisting coherence.
Magnetic fields further enable a dual-control strategy of dark coherence: (i) an out-of-plane component stabilizes coherence, and (ii) an in-plane component enables its optical readout, in well-defined and experimentally available field regimes.
These findings offer a versatile approach to accessing dark states via controlled coherence transfer, opening avenues for using bright-dark transitions in quantum information processing.

\begin{figure}
\centering
\includegraphics[width=0.9\linewidth]{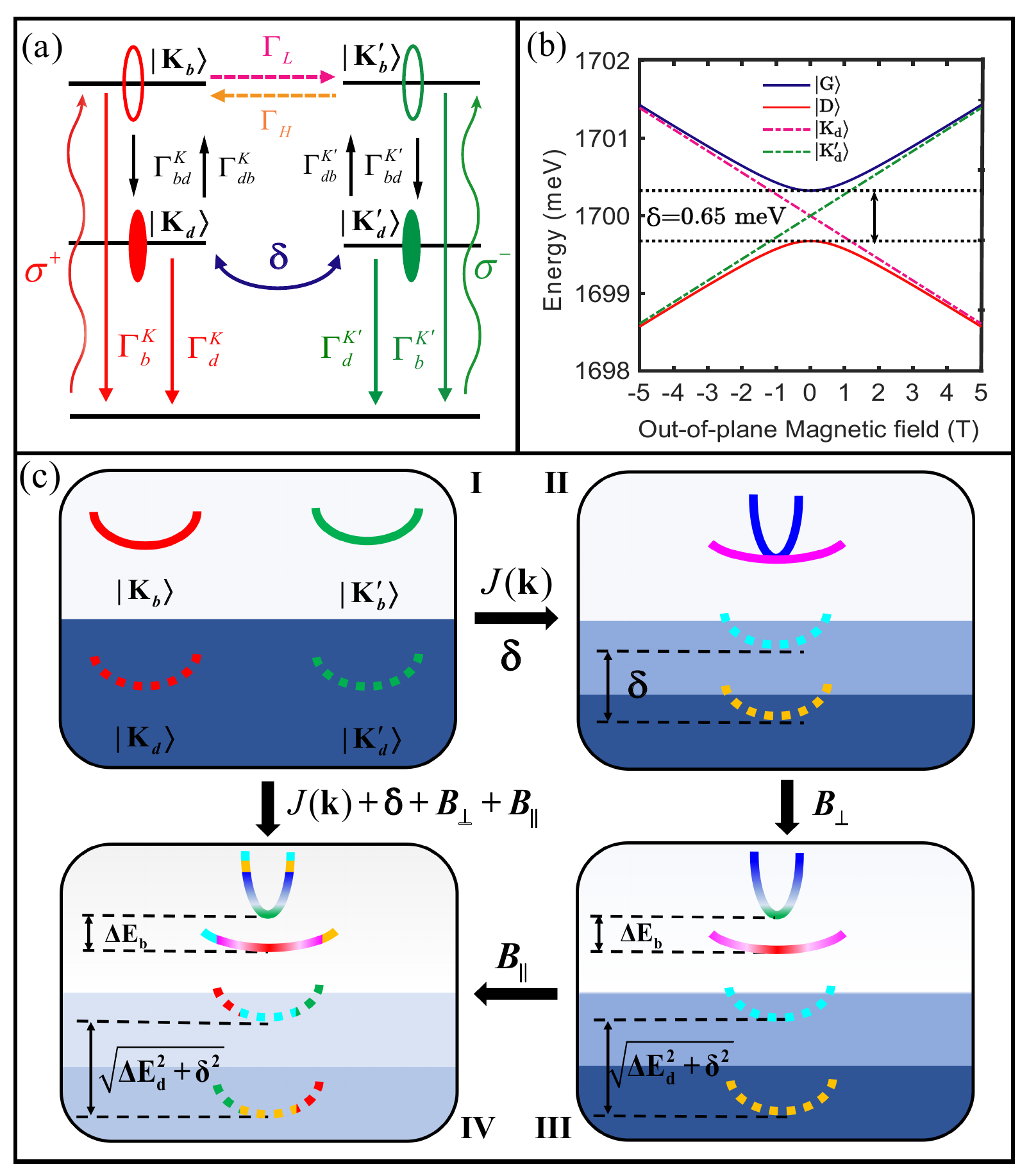}
\caption{(a) Five-level model with bright  and dark  excitons.
Coherent intervalley coupling in the dark manifold arises from SR exchange;
Incoherent processes originate from LR exchange and system-reservoir coupling, yielding Lindblad rates $\Gamma_i$:
$\Gamma_{b}^{K,K'}$ ($\Gamma_{d}^{K,K'}$) for bright (dark) recombination, $\Gamma_{H,L}$ for bright-exciton intervalley scattering, and $\Gamma_{bd}^{K(K')}$ for bright-dark scattering.
(b) Energies of  spin-unlike excitons $|\text{K}_d\rangle$, $|\text{K}'_d\rangle$  and exchange-dressed eigenstates $|\text{G}\rangle$ (grey), $|\text{D}\rangle$ (``truly'' dark) versus out-of-plane field.
(c) Exciton dispersions.
I: Spin-like and spin-unlike  valley excitons (red/green);
I$\rightarrow$II: LR and SR hybridize valely states into valley-mixed bright (blue/pink) and dark (cyan/orange) excitons;
II$\rightarrow$III: Out-of-plane field tunes valley mixing and splitting;
III$\rightarrow$IV: In-plane field  additionally mix bright and dark sectors.
Color gradients indicate momentum-dependent composition.
}
\label{figure1}
\end{figure}

\textit{Model Hamiltonian and valley dynamics}. 
We develop a microscopically-grounded open-quantum-system framework within a five-level model
comprising bright (spin-like) excitons $|\text{K}_b\rangle$; $|\text{K}'_b\rangle$), dark (spin-unlike) counterparts ($|\text{K}_d\rangle$; $|\text{K}'_d\rangle$)~\cite{footenotedark}, and the ground state ($|0\rangle$) [Fig.~\ref{figure1}(a)]~\cite{PhysRevLett.123.096803,PhysRevLett.129.067402}.
Unless stated otherwise, we use the term ``dark'' to denote spin-unlike excitons, reflecting their optical inactivity to in-plane polarized light~\cite{footenotedark}.
The total Hamiltonian is $\mathcal{H}=H_0+H_{\rm{E}}+H_{\rm{I}}$~\cite{PhysRevB.108.035419}: $H_0$ for valley  subsystem, $H_{\rm{E}}$ for environment reservoir, and  $H_{\rm{I}}$ for their interaction,
\begin{equation}
\label{eq1}
\begin{split}
&H_0=H_{\rm{ex}}+H_{{\rm SR}}+H_{{\rm LR}}+H_{B_\perp}+H_{B_\parallel},\\
  &H_{\rm{E}}=\hbar\omega_La_L^{\dag}a_L+\hbar\omega_Ra_R^{\dag}a_R+\textstyle\sum_{\xi}\hbar\omega_{\xi}a_{\xi}^{\dag}a_{\xi}\\
&\quad~+\textstyle\sum_{\mathbf{q},\xi}\hbar \tilde{\omega}_{\mathbf{q},\xi}b_{\mathbf{q},\xi}^{\dag}b_{\mathbf{q},\xi},~\xi=K,K',\\
&H_{\rm{I}}=(\Omega_K^b\sigma_{K-}^ba_R^{\dag}+\Omega_{K'}^b\sigma_{K'-}^ba_L^{\dag}+\rm{H.c.})\\
&\quad~+(\textstyle\sum_{\xi}\Omega_{\xi}^d\sigma_{\xi-}^da_{\xi}^{\dag}+\textstyle\sum_{\mathbf{q},\xi}\Omega_{\xi}^s
(\mathbf{q})\sigma_\xi^{db}b_\xi^\dag+\rm{H.c.}),
\end{split}
\end{equation}
where $a_L^{\dag}(a_L)$ and $a_R^{\dag}(a_R)$ are the creation (annihilation) operators for the left- and right-CP photons with frequencies $\omega_L$ and $\omega_R$, and $a_{\xi}^{\dag}(a_{\xi})$ for the out-of-plane polarized photon with frequency $\omega_\xi$; $b_{\mathbf{q},\xi}^{\dag}(b_{\mathbf{q},\xi})$ represent phonon operators with frequencies $\tilde {\omega}_{\mathbf{q},\xi}$.
And, $\Omega_{\xi}^b$ ($\Omega_{\xi}^d$) denotes the bright (dark) exciton-photon coupling governing recombination, and $\Omega_\xi^s$ represents the spin-orbit-coupling (SOC) mediated exciton-phonon coupling, which enables intravalley bright-dark scattering accompanied with phonon absorption or emission~\cite{sciadvabf3759,robert2020,Wang012106,PhysRevLett.115.257403}.
The lowering operators read $\sigma_{\xi-}^b=|0\rangle\langle\xi_b|$ and $\sigma_{\xi-}^d=|0\rangle\langle\xi_d|$, and $\sigma_{\xi}^{db}=|\xi_d\rangle\langle\xi_b|$.
The valley Hamiltonian $H_0$ comprises: {$H_{\rm{ex}}$} for exciton energies, \fu{$H_{\rm LR}$} for intervalley  long-range (LR) and $H_{\rm SR}$ for short-range (SR) electron-hole exchange interactions, and $H_{B_\perp}$ ($H_{B_\parallel}$) for out-of-plane (in-plane) magnetic effects,
\begin{equation}
\label{eq2}
\begin{split}
&H_{\rm{ex}}=\textstyle\sum_{\xi}(E_bc_{\xi b}^{\dag}c_{\xi b}+E_dc_{\xi d}^{\dag}c_{\xi d}),\\
  &H_{{\rm SR}}=\frac{\updelta}{2}c_{Kd}^{\dag}c_{K'd} + \rm{H.c.},\\
  &H_{\rm LR}=-J(\textbf{k})e^{-i2\theta}c_{Kb}^{\dag}c_{K'b}+\rm{H.c.},\\
&H_{B_\perp}=\beta_\perp\textstyle\sum_{\xi}(g_bc_{\xi b}^{\dag}c_{\xi b}+g_dc_{\xi d}^{\dag}c_{\xi d})\bar\sigma_z,\\
&H_{B_\parallel}=\beta_-g_\parallel(c_{Kb}^{\dag}c_{Kd}-c_{K'd}^{\dag}c_{K'b})+\rm{H.c.},
\end{split}
\end{equation}
with $c_{\xi b}^{\dag}$ ($c_{\xi d}^{\dag}$) the creation operator for bright (dark) excitons, and {$E_{b(d)}$ the corresponding excitonic energies with the \emph{intravalley} SR exchange energy (a blue shift of $\updelta$) being absorbed.
The \emph{intervalley} SR exchange coherently couples the spin-unlike states, yielding \emph{exchange-dressed} eigenstates $|\text{G}\rangle,|\text{D}\rangle=(|\text{K}_d\rangle\pm|\text{K}'_d\rangle)/\sqrt{2}$: a higher-energy grey state $|\text{G}\rangle$ that couples to out-of-plane polarization, and a lower-energy ``truly'' dark state $|\text{D}\rangle$ that is strictly optically inactive.
Their eigenenergies are obtained by diagonalizing the Hamiltonian of the $2\times2$ dark manifold [Fig.~\ref{figure1}(b)], yielding $E_{\text{G},\text{D}}=E_+\pm\sqrt{E_-^2+\updelta^2/4}$ with $E_\pm=(E_d^K\pm E_d^{K'})$.
A schematic distinction between spin-unlike exciton (prior to exchange)  and exchange-dressed eigenstates is provided in the Supplementary Material (SM) (Fig.~S1).
The LR exchange  acts as a momentum (\textbf{k})-dependent field whose random reorientation drives incoherent bright-exciton intervalley scattering~\cite{PhysRevB.47.15776,footnote-exch}. 
The Zeeman terms $\beta_{\perp}=\mu_BB_\perp/2$ (with $\bar\sigma_z=I\otimes\sigma_z$) lift valley degeneracy, while $\beta{\pm}=\mu_B(B_x\pm iB_y)/2$ induce coherent bright-dark mixing [Fig.~\ref{figure1}(c)]~\cite{PhysRevLett.123.096803,PhysRevB.104.195424,Qu2019}.

We then start from  Eqs.~\eqref{eq1} and \eqref{eq2} to derive the master equation determining the system dynamics~\cite{nanolett.8b01484,nl503799t,zeng2012valley,PhysRevB.97.115425,PhysRevB.101.085406, 10.1063/1.5112823,Baranowski2017}, accounting for unitary evolution ($\mathcal{L}_0$), intra- and intervalley scattering ($\mathcal{L}_f$) (SM; Secs.~\Rmnum{1} and \Rmnum{2}), radiative recombination ($\mathcal{L}_r$), and pure dephasing ($\mathcal{L}_p$),
\begin{equation}
\label{eq3}
\begin{split}
\frac{d}{dt}\rho(t)=\mathcal{L}_0\rho(t)+\mathcal{L}_f\rho(t)+\mathcal{L}_r\rho(t)+\mathcal{L}_p\rho(t),
\end{split}
\end{equation}
where the  Lindblad rates are illustrated in Fig.~\ref{figure1}(a), with $\Gamma_{H(L)}$ associated with $J(\mathbf{k})$ ($H_{\rm LR}$)~\cite{footnote-impurity} and $\Gamma_{bd}^{\rm K(K')}$ with $\Omega_\xi^s$ ($H_{\rm I}$).  
Full derivations of Eq.~\eqref{eq3} and the resulting correspondence between Hamiltonian terms [Eqs.~\eqref{eq1} and \eqref{eq2}] and scattering rates [$\Gamma_i$; Fig.~\ref{figure1}(a)],  see the SM (Sec.~\Rmnum{6}).

For direct experimental relevance, we quantify valley coherence using the $l_1$-norm $\mathcal{C}_{l=b,d}(\rho)=|\rho_l^{KK'}|+|\rho_l^{K'K}|$~\cite{PhysRevLett.113.140401}, which measures the magnitude of off-diagonal density-matrix elements,  and is  directly linked to the experimentally accessible degree of linear polarization~\cite{qiu2019room,PhysRevB.104.L121408}. Also, the associated coherence time $\uptau_l^{\mathcal{C}}$, defined as the characteristic decay time of $\mathcal{C}_l(\rho)$~\cite{PhysRevLett.121.116102,footenotetime}, can be extracted from homogeneous linewidths  measured using two-dimensional coherent spectroscopy~\cite{hao_moody2016}.

The initial state for considered polarized excitation field is $|\uppsi_0(\epsilon)\rangle=\sqrt{\frac{1-\epsilon}{2}}|\text{K}_b\rangle+e^{i\bar\varphi}\sqrt{\frac{1+\epsilon}{2}}|\text{K}'_b\rangle$, with $\bar\varphi$ the phase angle, $\epsilon=\tan[\frac{1}{2}\arcsin(\frac{2A_xA_y\sin2\chi}{A_x^2+A_y^2})]$ the ellipticity, and $2\chi$ the ellipticity angle [Fig.~\ref{figure2}(d)]~\cite{Chen20}.
At the equator ($\chi=0$) of Poincare sphere, ellipticity $\epsilon=0$ corresponds to LP excitation, with the polarization direction determined by the longitude angle $2\psi$.
The cases $\chi=\pm\pi/4$ respectively represent right- and left-handed CP excitations with ellipticity $\epsilon=\pm1$, while intermediate values $0<|\epsilon|<1$ corresponds to EP excitation.
The initial coherence hence can be characterized as $\mathcal{C}_0(\rho)=\sqrt{1-\epsilon^2}$.
Table~\ref{table1} lists central parameters for monolayer  WSe$_2$ in our calculations, including the $g$ factors and SR exchange parameter $\updelta$.
Values are benchmarked against magneto-optical spectroscopy~\cite{PhysRevLett.123.096803,PhysRevLett.126.067403,aivazianmagnetic2015,tepliakov2020} and first-principles calculations~\cite{PhysRevLett.124.226402,PhysRevB.101.235408,Faria2022}, with the scattering rates $\Gamma_i$ [Fig.~\ref{figure1}(a)] are further supported by time-resolved photoluminescence measurements~\cite{PhysRevB.90.075413,nanolett.8b01484,PhysRevB.96.155423}, together with theoretical calculations~\cite{nl503799t}.
\begin{table}[H]
   \caption{Material parameters used in the computation, including the $g$ factors and SR exchange parameter $\updelta$, as well as the ones used to
    determine recombination and scattering rates of $\Gamma_i$ [Fig.~\ref{figure1}(a)].}
  \begin{ruledtabular}
     \begin{tabular}{cccccccccc}
       			$g_b$& $g_d$& $g_\parallel$& $\updelta$& $\uptau_v$& $\alpha$& $\uptau_\xi^0$& $\Gamma_{bd}$& $\Gamma_d$  \\ \hline
       -4.2$^{\rm a}$  & -9.6$^{\rm a}$  & $2^{\rm a}$  & 0.65$^{\rm a}$  & 4$^{\rm b}$  &  $3.5\times10^5$$^{\rm b}$  & 0.22$^{\rm b}$  & 1/4$^{\rm b}$  & 1/110$^{\rm b}$  \\ \hline
		 /& /& /& meV& ps& meV$^{-3}$ & ps& ps$^{-1}$& ps$^{-1}$\\      		
\end{tabular}
  \end{ruledtabular}
    \begin{tablenotes}
\item[a]\raggedright From Ref.~\cite{PhysRevLett.123.096803,tepliakov2020,PhysRevLett.126.067403,aivazianmagnetic2015,PhysRevLett.124.226402,PhysRevB.101.235408,Faria2022}.
\item[b]\raggedright From Ref.~\cite{PhysRevB.90.075413,nl503799t,PhysRevB.96.155423,nanolett.8b01484,Baranowski2017}.
    \justifying Parameters determining the recombination and scattering rates $\Gamma_i$ in Fig.~\ref{figure1}(a):
 $\Gamma_{H,L}$ is set  by the zero-field scattering time $\uptau_v$ and exciton-phonon coupling strength $\alpha$;  $\Gamma_{b}^{\xi}$ by the zero-temperature radiative  lifetime $\uptau_\xi^0$; and $\Gamma_{bd}^{\xi}$ and $\Gamma_{d}^{\xi}$ are taken as constants $\Gamma_{bd}$ and $\Gamma_{d}$, respectively (SM; Sec.~\Rmnum{5}).
\end{tablenotes}
	\label{table1}
\end{table}

\textit{Ellipticity-engineered transition of valley coherence}.
By comparing LP [Fig.~\ref{figure2}(a)] and CP [Fig.~\ref{figure2}(b)] excitations, we find a coherence transition between bright (red line) and dark (blue line) states by switching the excitation mode.
Bright coherence only emerges under LP excitation [Fig.~\ref{figure2}(a), red lines], with a short coherence time of $\uptau_b^\mathcal{C}\approx$1 ps due to rapid decoherence [Fig.~\ref{figure2}(a), green circle].
The momentum-dependent LR exchange acts as a stochastic in-plane magnetic field that randomizes the valley pseudospin, leading to an exponential decay as $\mathcal{C}_b(\rho)=\mathcal{C}_0(\rho)e^{-\bar{\upgamma}_bt}$, with $\bar{\upgamma}_b=\Gamma_b+\upgamma_b$ the corresponding decay rate.
And, under CP excitation with $\mathcal{C}_0(\rho)=0$, no measurable coherence evolution occurs.

Remarkably, dark-exciton valley coherence can emerge spontaneously without any \emph{initial} coherent drive [Fig.~\ref{figure2}(b), blue line].
The CP excitation creates a valley population imbalance in the bright sector, which is transferred incoherently to the dark manifold via phonon-assisted bright-dark scattering enabled by SOC, carrying no phase information.
The intervalley SR exchange then coherently couples the imbalanced dark-valley populations, generating finite off-diagonal density-matrix elements and hence dark coherence. By contrast, the LP  excitation produces bright-exciton coherence but no dark-valley population imbalance, and therefore fails to generate dark coherence [Fig.~\ref{figure2}(b), red line]. Thus, dark-exciton coherence arises not from initial coherent drive,
but from the exchange-driven, population-imbalance-activated mechanism.
\begin{figure}
\includegraphics[width=\linewidth]{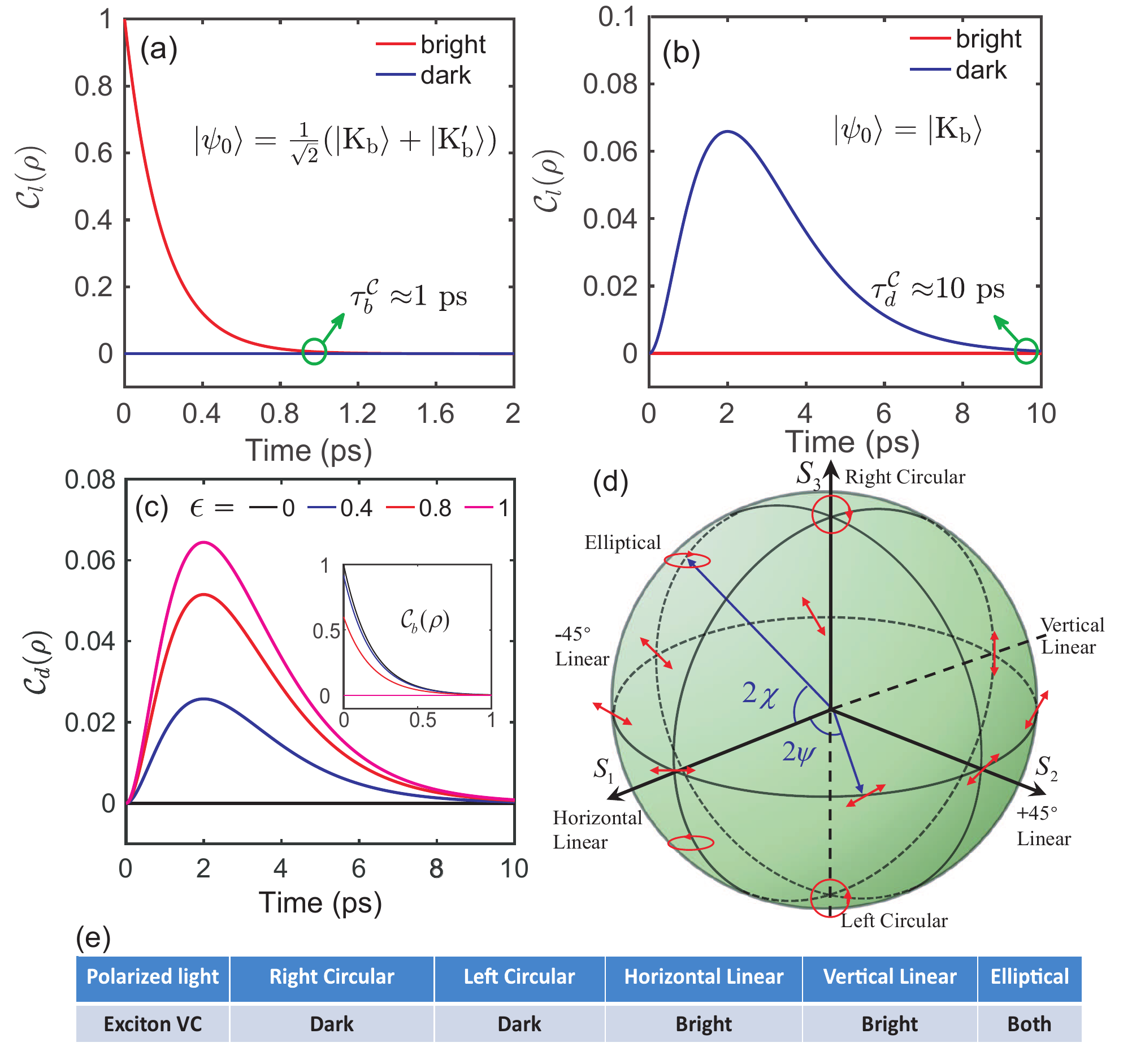}
\caption{Time evolutions of VC for bright [$\mathcal{C}_b(\rho)$: red line] and dark [$\mathcal{C}_d(\rho)$: blue line] excitons under LP (a), CP (b) and EP (c) excitations.
The green circles in (a) and (b) refer to the coherence times $\uptau_l^{\mathcal{C}}$.
(d) Poincare sphere schematic and optical polarization state representation.
Here $\chi$ and $\psi$ respectively represent the orientation angle of major axis for the polarization ellipse and the ellipticity angle.
(e) Summary of exciton coherence driven by different polarized lights.}
\label{figure2}
\end{figure}
We further quantify the selectivity in forming dark coherence via the analysis solution,
\begin{equation}
\label{eq4}
\begin{split}
\mathcal{C}_d(\rho)=|\updelta\Gamma_{bd}\upeta_b|\Bigg[\frac{e^{-\upgamma_{\rm{vp}}t}}{\Lambda}+
\sum_{i=1}^2\frac{(-1)^{i+1}e^{\kappa_it}}{(\kappa_1-\kappa_2)(\kappa_1+\upgamma_{\rm{vp}})}\Bigg],
\end{split}
\end{equation}
with $\kappa_{1,2}=-\frac{1}{2}\Big[(\bar\upgamma_d+\bar\Gamma_d)\mp\sqrt{(\bar\upgamma_d-
\bar\Gamma_d)^2-4\updelta^2}\Big]$, $\Lambda=\upgamma_{\rm{vp}}^2-(\bar\upgamma_d+\bar\Gamma_d)+\bar\upgamma_d\bar\Gamma_d+\updelta^2$,
see the SM \fu{(Sec.~\Rmnum{7})} for detailed derivation.
Equation~(\ref{eq4}) reveals three essential factors for establishing dark coherence: (i) the bright-to-dark exciton scattering rate ($\Gamma_{bd}$), (ii) intervalley SR exchange ($\updelta$), and (iii) initial bright-excitons population imbalance ($\upeta_b$).

For arbitrary polarized excitation, the ellipticity $\epsilon$ continuously tunes the initial coherence from maximal under LP ($\epsilon=0$) to zero for CP ($\epsilon=\pm1$), with EP ($0<|\epsilon|<1$) yielding intermediate coherence.
In Fig.~\ref{figure2}(c), dark coherence intensity grows markedly with ellipticity and peaks under CP excitation, while bright coherence declines--dropping by 50\% once $\epsilon\textgreater0.8$ (see the inset).
A summary of Ellipticity-dependent coherence types is provided in Fig.~\ref{figure2}(e).
In contrast to prior studies that treat bright and dark excitons as largely independent coherence platforms accessed via magnetic fields, including ``dark-exciton valley toolkit''~\cite{PhysRevLett.123.096803,PhysRevB.96.155423,zhang883}, momentum-space coherent superpositions~\cite{Xuezhi2022} and waveguide-assisted dark readout~\cite{Tang2019}, our results establish a new ellipticity-based dynamically control principle that links bright and dark manifolds, demonstrating ellipticity as a powerful control knob for valley coherent manipulation and identifying an exchange-driven, population-imbalance-activated mechanism that enables spontaneous dark coherence under CP excitation.

\begin{figure}
\includegraphics[width=\linewidth]{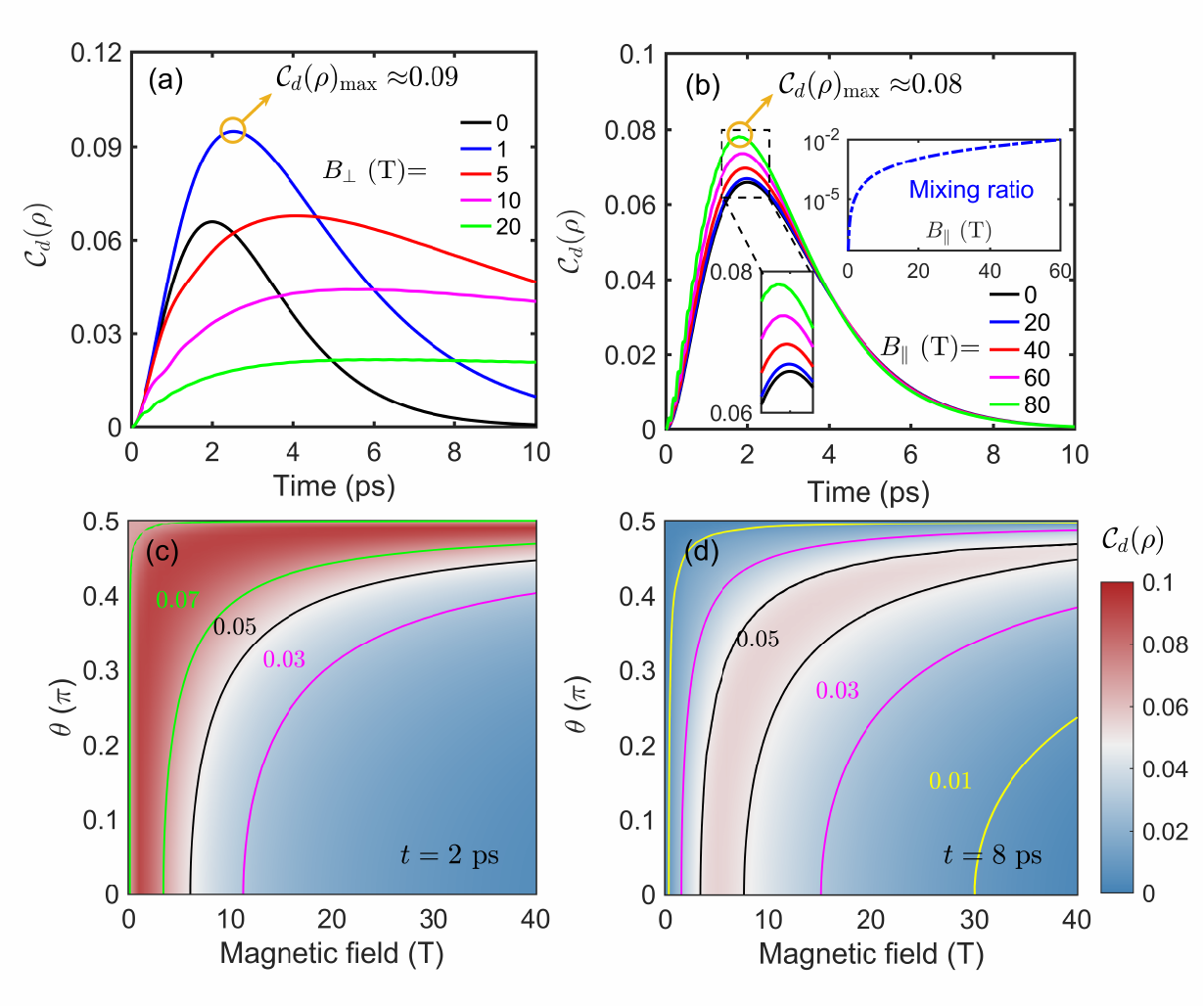}
\caption{The time evolutions of dark coherence for different out-of-plane (a) and in-plane (b) magnetic field strength.
The yellow circles in (a) and (b) refer to the maximum coherence intensity $\mathcal{C}_d(\rho)_{\max}$.
The inset in (b) shows the ratio of bright-dark mixing [Eq.~\eqref{eqcoherlu}].
The dark coherence at $t=2$ (c) and $8$ ps (d) as a function of magnetic field strength and azimuth angle $\theta(\pi)=\arctan(B_{\parallel}/B_{\perp})$.
Several values of contour lines of coherence intensity $\mathcal{C}_d(\rho)$ are also shown.}
\label{figure3}
\end{figure}
\textit{Magnetic enabled coherence manipulation and detection}.
Here we demonstrate magnetic control of spontaneous coherence.
An out-of-plane field governs coherent dynamics by competing with the SR exchange interaction.
We find a moderate field ($B_\perp\sim$1 T) suppresses exchange-induced decoherence, raising the coherence peak to 0.09 [Fig.~\ref{figure3}(a), yellow circle], although energetic splitting favors bare dark states over coherent superpositions.
Stronger fields ($B_\perp\sim$5--20 T) quench exchange-driven oscillations, extending coherence lifetime beyond 10 ps despite reduced intensity.
We also confirm delayed bright-exciton coherence decay via valley-split suppression of exchange interaction ({SM; Fig.~S2}).
All field strengths are well within magneto-optical experiments.

An in-plane magnetic field brightens dark excitons by transferring oscillator strength from bright states, yielding two key effects: (i) coherence enhancement at high fields, with the intensity of approximately 0.08 at $B_{\parallel}=80$ T [Fig.~\ref{figure3}(b), yellow circle]; and (ii) optical observability of dark coherence, which can be captured by the hybrid eigenstates in basis $\{|\text{K}_b\rangle,|\text{K}'_b\rangle,|\text{K}_d\rangle,|\text{K}'_d\rangle\}$~\cite{PhysRevLett.123.096803},
\begin{equation}
\label{eqcoherlu}
\begin{split}
|\rm{H}\rangle&=N\left[-\frac{\mu_Bg_{\parallel}B_{\parallel}}{2\Delta E_{bd}}\sin\varphi,-\frac{\mu_Bg_{\parallel}B_{\parallel}}{2\Delta E_{bd}}\cos\varphi,\sin\varphi,\cos\varphi\right],\\
|\text{L}\rangle&=N\left[-\frac{\mu_Bg_{\parallel}B_{\parallel}}{2\Delta E_{bd}}\cos\varphi,\frac{\mu_Bg_{\parallel}B_{\parallel}}{2\Delta E_{bd}}\sin\varphi,\cos\varphi,-\sin\varphi\right],
\end{split}
\end{equation}
with mixing angle $\varphi=\frac{1}{2}\arcsin\left[g_d\mu_BB_{\perp}/\sqrt{\updelta^2+(g_d\mu_BB_{\perp})^2}\right]+\frac{\pi}{4}$, normalization factor $N=1/\sqrt{1+(\mu_Bg_{\parallel}B_{\parallel}/2\Delta E_{bd})^2}$.
The corresponding eigenvalues are $E_{\rm{H,L}}=E_+\pm\sqrt{4E_-^2+\updelta^2}/2-\frac{(g_{\parallel}\mu_BB_{\parallel})^2}{4\Delta E_{bd}}$.
The optical visibility of these hybrid states is governed by the admixture of the bright component into the dark sector, which scales as $(N\mu_Bg_{\parallel}B_{\parallel}/2\Delta E_{bd})^2$ and is thus  tunable by the in-plane field [Fig.~\ref{figure3}(b), inset].
As $B_\parallel\rightarrow0$, clearly the original $|\text{G}\rangle$ and $|\text{D}\rangle$ states are restored~\cite{PhysRevLett.123.096803,Qu2019}.
\begin{figure}
\includegraphics[width=\linewidth]{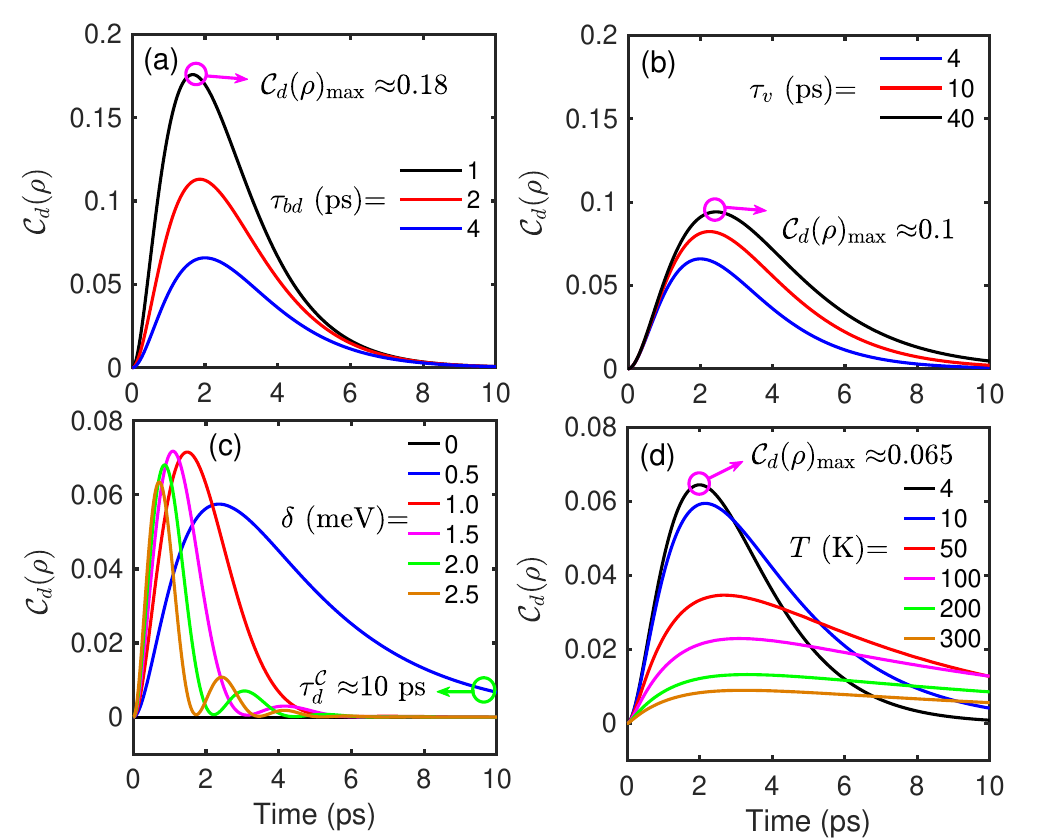}
\caption{The time evolution of dark coherence for different intravalley scattering times $\uptau_{bd}$ (a), intervalley scattering times $\uptau_{\text{v}}$ (b), exchange interactions $\updelta$ (c) and temperatures $T$ (d).
The pink circles in (a), (b) and (d) refer to the maximum coherence intensities $\mathcal{C}_d(\rho)_{\max}$, while the green circle in (c) refers to the coherence time $\uptau_d^{\mathcal{C}}$.}
\label{figure4}
\end{figure}

We also find that the response of dark coherence to the magnetic azimuthal angle $\theta$ evolves distinctly across different dynamical stages.
In the coherence growth phase [Fig.~\ref{figure3}(c)], the angular dependence is weak when $B\textless5$ T, while at higher fields, larger $\theta$ enhances intensity and mitigate field-induced suppression, yielding high coherence (green contours) in either low-$B$ or large-$\theta$ regimes.
In contrast, in the coherence decay phase [Fig.~\ref{figure3}(d)], large angles ($\theta\textgreater0.45\pi$) strongly degrade the coherence (yellow contours).
This stage-dependent behavior demonstrates how field strength and azimuthal angle jointly  enable temporally tailored coherence control.

\textit{The role of intrinsic dynamical quantities and thermal effects}.
For a comprehensive understanding, we examine how system parameters and temperature affect the dark-coherence dynamics, and identify two effective coherence enhancement pathways.
First, shortening the intravalley bright-dark scattering time $\uptau_{bd}$ to 1 ps boosts the coherence intensity to 0.18 (black line)--threefold increase over $\uptau_{bd}=4$ ps (blue line) [Fig.~\ref{figure4}(a)].
Further gains are anticipated from the direct generation of dark excitons, such as by using vortex beams to prepare the initial state $|\uppsi_0\rangle=|\text{K}_d\rangle$~\cite{acsnano.4c01881,PhysRevB.108.125435}.
Second, extending the intervalley scattering time $\uptau_{\text{v}}=40$ ps (black line) elevates the intensity to 0.1 (pink circle) and prolongs the coherence time to 10 ps [Fig.~\ref{figure4}(b)].
These results establish monolayers with rapid bright-dark scattering and slow intervalley scattering as an ideal platform for robust dark coherence.

Figure~\ref{figure4}(c) reveals that stronger exchange coupling accelerates decoherence, with peak intensity maximized at $\updelta=1.5$ meV (pink curve) and coherence time maximized at $\updelta=0.5$ meV (green circle).
This highlights a fundamental trade-off between coherent manipulation speed (oscillation frequency between two dark states) and coherence duration: higher oscillation frequencies intensify environmental coupling and reduce coherence time~\cite{RevModPhys.95.025003}.
Thermal effects are shown in Fig.~\ref{figure4}(d).
We find that lower temperatures suppress phonon assisted dephasing~\cite{moody2015intrinsic}, which enhances the coherence intensity, yielding a maximum value of $C_d(\rho)_{\rm max}\approx 0.065$ at 4 K. Notably, while elevated temperature degrades coherence intensity, it simultaneously prolongs the recombination lifetime of bright excitons~\cite{nl503799t}. Once the lifetime exceeds 10 ps, residual bright excitons continue feeding dark states via intravalley scattering, unexpectedly extending coherence time despite weakened overall coherence [cf. the black and red lines in Fig.~\ref{figure4}(d)]. 
We emphasize that here the counterintuitive behavior between coherence intensity and coherence lifetime arises from two competing temperature-dependent processes, i.e.,  enhanced phonon-induced dephasing (reducing coherence intensity) and suppressed radiative recombination with sustained population transfer (extending coherence lifetime) with increasing temperature.

\textit{Experimental detection scheme}--An hexagonal boron nitride (hBN)-encapsulated monolayer WSe$_2$ is excited by a pulsed laser with continuously adjustable polarization ellipticity, while an in-plane magnetic field is applied  to induce bright-dark mixing. Energy-resolved PL is collected with a spectrometer and analyzed using linear polarizers to separately measure the degree of LP of the bright-exciton peak and the magnetically brightened ``truly'' dark/grey exciton peaks~\cite{PhysRevLett.123.096803}. Bright-dark coherence transfer is identified by an ellipticity-controlled redistribution of LP, manifested as a decrease of polarization in the bright-exciton emission accompanied by the growth of polarized emission from the dark sectors.
Time-resolved Kerr-rotation measurements~\cite{PhysRevB.90.161302,nanolett01727} can further corroborate the transfer by revealing a long-lived polarization signal at the dark sector, exceeding the bright-exciton radiative lifetime.

\textit{Concluding remarks}--We have developed a unified, microscopically-grounded open-quantum-system framework that treats optical driving, SOC-enabled bright-dark scattering, LR and SR exchange interaction, dissipation, and magnetic-field control on equal footing. 
Then, we establish a distinct dynamical control principle in which excitation ellipticity continuously redistributes valley coherence between bright and dark manifolds; and  identify an exchange-driven, population-imbalance-activated mechanism that enables the spontaneous emergence of dark coherence under CP  excitation, without any initial coherence.
These provide a versatile toolbox for controlling dark coherence in practical implementations, and establish a new pathway for harnessing valley-coherence transitions between bright and dark excitons in future quantum control technologies.
As a remark, the chiral-phonon-mediated recombination of dark excitons reported by Liu et al.~\cite{PhysRevResearch.1.032007}, offers a complementary pathway
for phonon-assisted optical readout and control of dark excitons.  
More work is needed to explore this intereting possiblity.

\textit{Acknowledgments}. This work was supported by the National Natural Science Foundation of China (Grants No. 1250040826, No. 12274256, No. 11874236, No. 12022413, No. 11674331, No. 61674096, and No. 12575024), the Major Basic Program of Natural Science Foundation of Shandong Province (Grant No. ZR2021ZD01), the Natural Science Foundation of Shandong Province (Grant No. ZR2025QC1495, No. ZR2023LLZ015, and No. ZR2024LLZ012), the National Key R\&D Program of China (Grant No. 2022YFA1403200), and the “Strategic Priority Research Program (B)” of the Chinese Academy of Sciences (Grant No. XDB33030100).


\begin{thebibliography}{67}%
\makeatletter
\providecommand \@ifxundefined [1]{%
 \@ifx{#1\undefined}
}%
\providecommand \@ifnum [1]{%
 \ifnum #1\expandafter \@firstoftwo
 \else \expandafter \@secondoftwo
 \fi
}%
\providecommand \@ifx [1]{%
 \ifx #1\expandafter \@firstoftwo
 \else \expandafter \@secondoftwo
 \fi
}%
\providecommand \natexlab [1]{#1}%
\providecommand \enquote  [1]{``#1''}%
\providecommand \bibnamefont  [1]{#1}%
\providecommand \bibfnamefont [1]{#1}%
\providecommand \citenamefont [1]{#1}%
\providecommand \href@noop [0]{\@secondoftwo}%
\providecommand \href [0]{\begingroup \@sanitize@url \@href}%
\providecommand \@href[1]{\@@startlink{#1}\@@href}%
\providecommand \@@href[1]{\endgroup#1\@@endlink}%
\providecommand \@sanitize@url [0]{\catcode `\\12\catcode `\$12\catcode
  `\&12\catcode `\#12\catcode `\^12\catcode `\_12\catcode `\%12\relax}%
\providecommand \@@startlink[1]{}%
\providecommand \@@endlink[0]{}%
\providecommand \url  [0]{\begingroup\@sanitize@url \@url }%
\providecommand \@url [1]{\endgroup\@href {#1}{\urlprefix }}%
\providecommand \urlprefix  [0]{URL }%
\providecommand \Eprint [0]{\href }%
\providecommand \doibase [0]{https://doi.org/}%
\providecommand \selectlanguage [0]{\@gobble}%
\providecommand \bibinfo  [0]{\@secondoftwo}%
\providecommand \bibfield  [0]{\@secondoftwo}%
\providecommand \translation [1]{[#1]}%
\providecommand \BibitemOpen [0]{}%
\providecommand \bibitemStop [0]{}%
\providecommand \bibitemNoStop [0]{.\EOS\space}%
\providecommand \EOS [0]{\spacefactor3000\relax}%
\providecommand \BibitemShut  [1]{\csname bibitem#1\endcsname}%
\let\auto@bib@innerbib\@empty
\bibitem [{\citenamefont {Xiao}\ \emph {et~al.}(2012)\citenamefont {Xiao},
  \citenamefont {Liu}, \citenamefont {Feng}, \citenamefont {Xu},\ and\
  \citenamefont {Yao}}]{PhysRevLett.108.196802}%
  \BibitemOpen
  \bibfield  {author} {\bibinfo {author} {\bibfnamefont {D.}~\bibnamefont
  {Xiao}}, \bibinfo {author} {\bibfnamefont {G.-B.}\ \bibnamefont {Liu}},
  \bibinfo {author} {\bibfnamefont {W.}~\bibnamefont {Feng}}, \bibinfo {author}
  {\bibfnamefont {X.}~\bibnamefont {Xu}},\ and\ \bibinfo {author}
  {\bibfnamefont {W.}~\bibnamefont {Yao}},\ }\bibfield  {title} {\bibinfo
  {title} {Coupled spin and valley physics in monolayers of {MoS}$_{2}$ and
  other group-{VI} dichalcogenides},\ }\href
  {https://doi.org/10.1103/PhysRevLett.108.196802} {\bibfield  {journal}
  {\bibinfo  {journal} {Phys. Rev. Lett.}\ }\textbf {\bibinfo {volume} {108}},\
  \bibinfo {pages} {196802} (\bibinfo {year} {2012})}\BibitemShut {NoStop}%
\bibitem [{\citenamefont {Ye}\ \emph {et~al.}(2014)\citenamefont {Ye},
  \citenamefont {Cao}, \citenamefont {O’Brien}, \citenamefont {Zhu},
  \citenamefont {Yin}, \citenamefont {Wang}, \citenamefont {Louie},\ and\
  \citenamefont {Zhang}}]{Ye2014}%
  \BibitemOpen
  \bibfield  {author} {\bibinfo {author} {\bibfnamefont {Z.}~\bibnamefont
  {Ye}}, \bibinfo {author} {\bibfnamefont {T.}~\bibnamefont {Cao}}, \bibinfo
  {author} {\bibfnamefont {K.}~\bibnamefont {O’Brien}}, \bibinfo {author}
  {\bibfnamefont {H.}~\bibnamefont {Zhu}}, \bibinfo {author} {\bibfnamefont
  {X.}~\bibnamefont {Yin}}, \bibinfo {author} {\bibfnamefont {Y.}~\bibnamefont
  {Wang}}, \bibinfo {author} {\bibfnamefont {S.~G.}\ \bibnamefont {Louie}},\
  and\ \bibinfo {author} {\bibfnamefont {X.}~\bibnamefont {Zhang}},\ }\bibfield
   {title} {\bibinfo {title} {Probing excitonic dark states in single-layer
  tungsten disulphide},\ }\href {https://doi.org/10.1038/nature13734}
  {\bibfield  {journal} {\bibinfo  {journal} {Nature}\ }\textbf {\bibinfo
  {volume} {513}},\ \bibinfo {pages} {214} (\bibinfo {year}
  {2014})}\BibitemShut {NoStop}%
\bibitem [{\citenamefont {Yao}\ \emph {et~al.}(2008)\citenamefont {Yao},
  \citenamefont {Xiao},\ and\ \citenamefont {Niu}}]{PhysRevB.77.235406}%
  \BibitemOpen
  \bibfield  {author} {\bibinfo {author} {\bibfnamefont {W.}~\bibnamefont
  {Yao}}, \bibinfo {author} {\bibfnamefont {D.}~\bibnamefont {Xiao}},\ and\
  \bibinfo {author} {\bibfnamefont {Q.}~\bibnamefont {Niu}},\ }\bibfield
  {title} {\bibinfo {title} {Valley-dependent optoelectronics from inversion
  symmetry breaking},\ }\href {https://doi.org/10.1103/PhysRevB.77.235406}
  {\bibfield  {journal} {\bibinfo  {journal} {Phys. Rev. B}\ }\textbf {\bibinfo
  {volume} {77}},\ \bibinfo {pages} {235406} (\bibinfo {year}
  {2008})}\BibitemShut {NoStop}%
\bibitem [{\citenamefont {Wang}\ \emph {et~al.}(2012)\citenamefont {Wang},
  \citenamefont {Kalantar-Zadeh}, \citenamefont {Kis}, \citenamefont
  {Coleman},\ and\ \citenamefont {Strano}}]{wangkalantar2012}%
  \BibitemOpen
  \bibfield  {author} {\bibinfo {author} {\bibfnamefont {Q.~H.}\ \bibnamefont
  {Wang}}, \bibinfo {author} {\bibfnamefont {K.}~\bibnamefont
  {Kalantar-Zadeh}}, \bibinfo {author} {\bibfnamefont {A.}~\bibnamefont {Kis}},
  \bibinfo {author} {\bibfnamefont {J.~N.}\ \bibnamefont {Coleman}},\ and\
  \bibinfo {author} {\bibfnamefont {M.~S.}\ \bibnamefont {Strano}},\ }\bibfield
   {title} {\bibinfo {title} {Electronics and optoelectronics of
  two-dimensional transition metal dichalcogenides},\ }\href
  {https://doi.org/10.1038/nnano.2012.193} {\bibfield  {journal} {\bibinfo
  {journal} {Nat. Nanotechnol.}\ }\textbf {\bibinfo {volume} {7}},\ \bibinfo
  {pages} {699–712} (\bibinfo {year} {2012})}\BibitemShut {NoStop}%
\bibitem [{\citenamefont {Kim}\ \emph {et~al.}(2025)\citenamefont {Kim},
  \citenamefont {Lee}, \citenamefont {Kim}, \citenamefont {Park}, \citenamefont
  {Kim}, \citenamefont {Choi}, \citenamefont {Watanabe}, \citenamefont
  {Taniguchi}, \citenamefont {Jo},\ and\ \citenamefont {Choi}}]{2whm-6xyv}%
  \BibitemOpen
  \bibfield  {author} {\bibinfo {author} {\bibfnamefont {H.}~\bibnamefont
  {Kim}}, \bibinfo {author} {\bibfnamefont {G.}~\bibnamefont {Lee}}, \bibinfo
  {author} {\bibfnamefont {J.}~\bibnamefont {Kim}}, \bibinfo {author}
  {\bibfnamefont {J.}~\bibnamefont {Park}}, \bibinfo {author} {\bibfnamefont
  {A.~S.}\ \bibnamefont {Kim}}, \bibinfo {author} {\bibfnamefont
  {J.}~\bibnamefont {Choi}}, \bibinfo {author} {\bibfnamefont {K.}~\bibnamefont
  {Watanabe}}, \bibinfo {author} {\bibfnamefont {T.}~\bibnamefont {Taniguchi}},
  \bibinfo {author} {\bibfnamefont {M.-H.}\ \bibnamefont {Jo}},\ and\ \bibinfo
  {author} {\bibfnamefont {H.}~\bibnamefont {Choi}},\ }\bibfield  {title}
  {\bibinfo {title} {Exciton dynamics in marginally twisted
  $\mathrm{WS}{\mathrm{e}}_{2}$ homobilayer: Role of interlayer coupling,
  phonons, and intervalley scattering},\ }\href
  {https://doi.org/10.1103/2whm-6xyv} {\bibfield  {journal} {\bibinfo
  {journal} {Phys. Rev. B}\ }\textbf {\bibinfo {volume} {112}},\ \bibinfo
  {pages} {104305} (\bibinfo {year} {2025})}\BibitemShut {NoStop}%
\bibitem [{\citenamefont {Xiang}\ \emph {et~al.}(2025)\citenamefont {Xiang},
  \citenamefont {Shinokita}, \citenamefont {Watanabe}, \citenamefont
  {Taniguchi},\ and\ \citenamefont {Matsuda}}]{sciadv.adr5562}%
  \BibitemOpen
  \bibfield  {author} {\bibinfo {author} {\bibfnamefont {Y.}~\bibnamefont
  {Xiang}}, \bibinfo {author} {\bibfnamefont {K.}~\bibnamefont {Shinokita}},
  \bibinfo {author} {\bibfnamefont {K.}~\bibnamefont {Watanabe}}, \bibinfo
  {author} {\bibfnamefont {T.}~\bibnamefont {Taniguchi}},\ and\ \bibinfo
  {author} {\bibfnamefont {K.}~\bibnamefont {Matsuda}},\ }\bibfield  {title}
  {\bibinfo {title} {Magnetic brightening and its dynamics of defect-localized
  exciton emission in monolayer two-dimensional semiconductor},\ }\href
  {https://doi.org/10.1126/sciadv.adr5562} {\bibfield  {journal} {\bibinfo
  {journal} {Sci. Adv.}\ }\textbf {\bibinfo {volume} {11}},\ \bibinfo {pages}
  {eadr5562} (\bibinfo {year} {2025})}\BibitemShut {NoStop}%
\bibitem [{\citenamefont {Chen}\ \emph {et~al.}(2018)\citenamefont {Chen},
  \citenamefont {Goldstein}, \citenamefont {Tong}, \citenamefont {Taniguchi},
  \citenamefont {Watanabe},\ and\ \citenamefont
  {Yan}}]{PhysRevLett.120.046402}%
  \BibitemOpen
  \bibfield  {author} {\bibinfo {author} {\bibfnamefont {S.-Y.}\ \bibnamefont
  {Chen}}, \bibinfo {author} {\bibfnamefont {T.}~\bibnamefont {Goldstein}},
  \bibinfo {author} {\bibfnamefont {J.}~\bibnamefont {Tong}}, \bibinfo {author}
  {\bibfnamefont {T.}~\bibnamefont {Taniguchi}}, \bibinfo {author}
  {\bibfnamefont {K.}~\bibnamefont {Watanabe}},\ and\ \bibinfo {author}
  {\bibfnamefont {J.}~\bibnamefont {Yan}},\ }\bibfield  {title} {\bibinfo
  {title} {Superior valley polarization and coherence of $2s$ excitons in
  monolayer {WSe}$_2$},\ }\href
  {https://doi.org/10.1103/PhysRevLett.120.046402} {\bibfield  {journal}
  {\bibinfo  {journal} {Phys. Rev. Lett.}\ }\textbf {\bibinfo {volume} {120}},\
  \bibinfo {pages} {046402} (\bibinfo {year} {2018})}\BibitemShut {NoStop}%
\bibitem [{\citenamefont {El~Houri}\ \emph {et~al.}(2024)\citenamefont
  {El~Houri}, \citenamefont {Khribach}, \citenamefont {Adnane}, \citenamefont
  {Moqine}, \citenamefont {Houça}, \citenamefont {Kamal},\ and\ \citenamefont
  {Belouad}}]{ElHouri2024}%
  \BibitemOpen
  \bibfield  {author} {\bibinfo {author} {\bibfnamefont {A.}~\bibnamefont
  {El~Houri}}, \bibinfo {author} {\bibfnamefont {A.}~\bibnamefont {Khribach}},
  \bibinfo {author} {\bibfnamefont {B.}~\bibnamefont {Adnane}}, \bibinfo
  {author} {\bibfnamefont {Y.}~\bibnamefont {Moqine}}, \bibinfo {author}
  {\bibfnamefont {R.}~\bibnamefont {Houça}}, \bibinfo {author} {\bibfnamefont
  {A.}~\bibnamefont {Kamal}},\ and\ \bibinfo {author} {\bibfnamefont
  {A.}~\bibnamefont {Belouad}},\ }\bibfield  {title} {\bibinfo {title} {Thermal
  quantum coherence: a comparative study of molybdenum disulfide versus
  graphene},\ }\href {https://doi.org/10.1088/1402-4896/ad91ec} {\bibfield
  {journal} {\bibinfo  {journal} {Phys. Scr.}\ }\textbf {\bibinfo {volume}
  {99}},\ \bibinfo {pages} {125119} (\bibinfo {year} {2024})}\BibitemShut
  {NoStop}%
\bibitem [{\citenamefont {Zhang}\ \emph {et~al.}(2025)\citenamefont {Zhang},
  \citenamefont {Hu}, \citenamefont {Perfetto},\ and\ \citenamefont
  {Stefanucci}}]{9spx-3hrq}%
  \BibitemOpen
  \bibfield  {author} {\bibinfo {author} {\bibfnamefont {Z.}~\bibnamefont
  {Zhang}}, \bibinfo {author} {\bibfnamefont {W.}~\bibnamefont {Hu}}, \bibinfo
  {author} {\bibfnamefont {E.}~\bibnamefont {Perfetto}},\ and\ \bibinfo
  {author} {\bibfnamefont {G.}~\bibnamefont {Stefanucci}},\ }\bibfield  {title}
  {\bibinfo {title} {Long-lived coherence between incoherent excitons revealed
  by time-resolved angle-resolved photoemission spectroscopy: An exact
  solution},\ }\href {https://doi.org/10.1103/9spx-3hrq} {\bibfield  {journal}
  {\bibinfo  {journal} {Phys. Rev. B}\ }\textbf {\bibinfo {volume} {111}},\
  \bibinfo {pages} {235124} (\bibinfo {year} {2025})}\BibitemShut {NoStop}%
\bibitem [{\citenamefont {Lan}\ \emph {et~al.}(2023)\citenamefont {Lan},
  \citenamefont {Xie}, \citenamefont {Fu},\ and\ \citenamefont
  {Qu}}]{PhysRevB.108.035419}%
  \BibitemOpen
  \bibfield  {author} {\bibinfo {author} {\bibfnamefont {K.}~\bibnamefont
  {Lan}}, \bibinfo {author} {\bibfnamefont {S.}~\bibnamefont {Xie}}, \bibinfo
  {author} {\bibfnamefont {J.}~\bibnamefont {Fu}},\ and\ \bibinfo {author}
  {\bibfnamefont {F.}~\bibnamefont {Qu}},\ }\bibfield  {title} {\bibinfo
  {title} {Magnetically tunable exciton valley coherence in monolayer
  {WS}$_{2}$ mediated by the electron-hole exchange and exciton-phonon
  interactions},\ }\href {https://doi.org/10.1103/PhysRevB.108.035419}
  {\bibfield  {journal} {\bibinfo  {journal} {Phys. Rev. B}\ }\textbf {\bibinfo
  {volume} {108}},\ \bibinfo {pages} {035419} (\bibinfo {year}
  {2023})}\BibitemShut {NoStop}%
\bibitem [{\citenamefont {Selig}\ \emph {et~al.}(2016)\citenamefont {Selig},
  \citenamefont {Bergh{\"a}user}, \citenamefont {Raja}, \citenamefont {Nagler},
  \citenamefont {Sch{\"u}ller}, \citenamefont {Heinz}, \citenamefont {Korn},
  \citenamefont {Chernikov}, \citenamefont {Malic},\ and\ \citenamefont
  {Knorr}}]{selig2016excitonic}%
  \BibitemOpen
  \bibfield  {author} {\bibinfo {author} {\bibfnamefont {M.}~\bibnamefont
  {Selig}}, \bibinfo {author} {\bibfnamefont {G.}~\bibnamefont
  {Bergh{\"a}user}}, \bibinfo {author} {\bibfnamefont {A.}~\bibnamefont
  {Raja}}, \bibinfo {author} {\bibfnamefont {P.}~\bibnamefont {Nagler}},
  \bibinfo {author} {\bibfnamefont {C.}~\bibnamefont {Sch{\"u}ller}}, \bibinfo
  {author} {\bibfnamefont {T.~F.}\ \bibnamefont {Heinz}}, \bibinfo {author}
  {\bibfnamefont {T.}~\bibnamefont {Korn}}, \bibinfo {author} {\bibfnamefont
  {A.}~\bibnamefont {Chernikov}}, \bibinfo {author} {\bibfnamefont
  {E.}~\bibnamefont {Malic}},\ and\ \bibinfo {author} {\bibfnamefont
  {A.}~\bibnamefont {Knorr}},\ }\bibfield  {title} {\bibinfo {title} {Excitonic
  linewidth and coherence lifetime in monolayer transition metal
  dichalcogenides},\ }\href {https://doi.org/10.1038/ncomms13279} {\bibfield
  {journal} {\bibinfo  {journal} {Nat. Commun.}\ }\textbf {\bibinfo {volume}
  {7}},\ \bibinfo {pages} {13279} (\bibinfo {year} {2016})}\BibitemShut
  {NoStop}%
\bibitem [{\citenamefont {Tao}\ \emph {et~al.}(2024)\citenamefont {Tao},
  \citenamefont {Shen}, \citenamefont {Jiang}, \citenamefont {Li},
  \citenamefont {Li}, \citenamefont {Ma}, \citenamefont {Zhao}, \citenamefont
  {Hu}, \citenamefont {Pistunova}, \citenamefont {Watanabe}, \citenamefont
  {Taniguchi}, \citenamefont {Heinz}, \citenamefont {Mak},\ and\ \citenamefont
  {Shan}}]{PhysRevX.14.011004}%
  \BibitemOpen
  \bibfield  {author} {\bibinfo {author} {\bibfnamefont {Z.}~\bibnamefont
  {Tao}}, \bibinfo {author} {\bibfnamefont {B.}~\bibnamefont {Shen}}, \bibinfo
  {author} {\bibfnamefont {S.}~\bibnamefont {Jiang}}, \bibinfo {author}
  {\bibfnamefont {T.}~\bibnamefont {Li}}, \bibinfo {author} {\bibfnamefont
  {L.}~\bibnamefont {Li}}, \bibinfo {author} {\bibfnamefont {L.}~\bibnamefont
  {Ma}}, \bibinfo {author} {\bibfnamefont {W.}~\bibnamefont {Zhao}}, \bibinfo
  {author} {\bibfnamefont {J.}~\bibnamefont {Hu}}, \bibinfo {author}
  {\bibfnamefont {K.}~\bibnamefont {Pistunova}}, \bibinfo {author}
  {\bibfnamefont {K.}~\bibnamefont {Watanabe}}, \bibinfo {author}
  {\bibfnamefont {T.}~\bibnamefont {Taniguchi}}, \bibinfo {author}
  {\bibfnamefont {T.~F.}\ \bibnamefont {Heinz}}, \bibinfo {author}
  {\bibfnamefont {K.~F.}\ \bibnamefont {Mak}},\ and\ \bibinfo {author}
  {\bibfnamefont {J.}~\bibnamefont {Shan}},\ }\bibfield  {title} {\bibinfo
  {title} {Valley-coherent quantum anomalous hall state in ab-stacked
  {MoTe}$_{2}$/{WSe}$_{2}$ bilayers},\ }\href
  {https://doi.org/10.1103/PhysRevX.14.011004} {\bibfield  {journal} {\bibinfo
  {journal} {Phys. Rev. X}\ }\textbf {\bibinfo {volume} {14}},\ \bibinfo
  {pages} {011004} (\bibinfo {year} {2024})}\BibitemShut {NoStop}%
\bibitem [{\citenamefont {Luo}\ \emph {et~al.}(2023)\citenamefont {Luo},
  \citenamefont {Whetten}, \citenamefont {Kravtsov}, \citenamefont {Singh},
  \citenamefont {Yang}, \citenamefont {Huang}, \citenamefont {Cheng},
  \citenamefont {Jiang}, \citenamefont {Belyanin},\ and\ \citenamefont
  {Raschke}}]{nanolett.2c04536}%
  \BibitemOpen
  \bibfield  {author} {\bibinfo {author} {\bibfnamefont {W.}~\bibnamefont
  {Luo}}, \bibinfo {author} {\bibfnamefont {B.~G.}\ \bibnamefont {Whetten}},
  \bibinfo {author} {\bibfnamefont {V.}~\bibnamefont {Kravtsov}}, \bibinfo
  {author} {\bibfnamefont {A.}~\bibnamefont {Singh}}, \bibinfo {author}
  {\bibfnamefont {Y.}~\bibnamefont {Yang}}, \bibinfo {author} {\bibfnamefont
  {D.}~\bibnamefont {Huang}}, \bibinfo {author} {\bibfnamefont
  {X.}~\bibnamefont {Cheng}}, \bibinfo {author} {\bibfnamefont
  {T.}~\bibnamefont {Jiang}}, \bibinfo {author} {\bibfnamefont
  {A.}~\bibnamefont {Belyanin}},\ and\ \bibinfo {author} {\bibfnamefont
  {M.~B.}\ \bibnamefont {Raschke}},\ }\bibfield  {title} {\bibinfo {title}
  {Ultrafast nanoimaging of electronic coherence of monolayer wse2},\ }\href
  {https://doi.org/10.1021/acs.nanolett.2c04536} {\bibfield  {journal}
  {\bibinfo  {journal} {Nano Lett.}\ }\textbf {\bibinfo {volume} {23}},\
  \bibinfo {pages} {1767} (\bibinfo {year} {2023})}\BibitemShut {NoStop}%
\bibitem [{\citenamefont {Xie}\ \emph {et~al.}(2024)\citenamefont {Xie},
  \citenamefont {Wu}, \citenamefont {Ding}, \citenamefont {Li}, \citenamefont
  {Chen}, \citenamefont {He}, \citenamefont {Liu}, \citenamefont {Wang},\ and\
  \citenamefont {Liu}}]{nanolett.4c01327}%
  \BibitemOpen
  \bibfield  {author} {\bibinfo {author} {\bibfnamefont {X.}~\bibnamefont
  {Xie}}, \bibinfo {author} {\bibfnamefont {B.}~\bibnamefont {Wu}}, \bibinfo
  {author} {\bibfnamefont {J.}~\bibnamefont {Ding}}, \bibinfo {author}
  {\bibfnamefont {S.}~\bibnamefont {Li}}, \bibinfo {author} {\bibfnamefont
  {J.}~\bibnamefont {Chen}}, \bibinfo {author} {\bibfnamefont {J.}~\bibnamefont
  {He}}, \bibinfo {author} {\bibfnamefont {Z.}~\bibnamefont {Liu}}, \bibinfo
  {author} {\bibfnamefont {J.-T.}\ \bibnamefont {Wang}},\ and\ \bibinfo
  {author} {\bibfnamefont {Y.}~\bibnamefont {Liu}},\ }\bibfield  {title}
  {\bibinfo {title} {Emergence of optical anisotropy in moiré superlattice via
  heterointerface engineering},\ }\href
  {https://doi.org/10.1021/acs.nanolett.4c01327} {\bibfield  {journal}
  {\bibinfo  {journal} {Nano Lett.}\ }\textbf {\bibinfo {volume} {24}},\
  \bibinfo {pages} {9186} (\bibinfo {year} {2024})}\BibitemShut {NoStop}%
\bibitem [{\citenamefont {Wu}\ \emph {et~al.}(2021)\citenamefont {Wu},
  \citenamefont {Taniguchi}, \citenamefont {Watanabe},\ and\ \citenamefont
  {Yan}}]{PhysRevB.104.L121408}%
  \BibitemOpen
  \bibfield  {author} {\bibinfo {author} {\bibfnamefont {Y.-C.}\ \bibnamefont
  {Wu}}, \bibinfo {author} {\bibfnamefont {T.}~\bibnamefont {Taniguchi}},
  \bibinfo {author} {\bibfnamefont {K.}~\bibnamefont {Watanabe}},\ and\
  \bibinfo {author} {\bibfnamefont {J.}~\bibnamefont {Yan}},\ }\bibfield
  {title} {\bibinfo {title} {Enhancement of exciton valley polarization in
  monolayer {MoS}$_{2}$ induced by scattering},\ }\href
  {https://doi.org/10.1103/PhysRevB.104.L121408} {\bibfield  {journal}
  {\bibinfo  {journal} {Phys. Rev. B}\ }\textbf {\bibinfo {volume} {104}},\
  \bibinfo {pages} {L121408} (\bibinfo {year} {2021})}\BibitemShut {NoStop}%
\bibitem [{\citenamefont {Jones}\ \emph {et~al.}(2013)\citenamefont {Jones},
  \citenamefont {Yu}, \citenamefont {Ghimire}, \citenamefont {Wu},
  \citenamefont {Aivazian}, \citenamefont {Ross}, \citenamefont {Zhao},
  \citenamefont {Yan}, \citenamefont {Mandrus}, \citenamefont {Xiao},\ and\
  \citenamefont {et~al.}}]{jones_yu2013}%
  \BibitemOpen
  \bibfield  {author} {\bibinfo {author} {\bibfnamefont {A.~M.}\ \bibnamefont
  {Jones}}, \bibinfo {author} {\bibfnamefont {H.}~\bibnamefont {Yu}}, \bibinfo
  {author} {\bibfnamefont {N.~J.}\ \bibnamefont {Ghimire}}, \bibinfo {author}
  {\bibfnamefont {S.}~\bibnamefont {Wu}}, \bibinfo {author} {\bibfnamefont
  {G.}~\bibnamefont {Aivazian}}, \bibinfo {author} {\bibfnamefont {J.~S.}\
  \bibnamefont {Ross}}, \bibinfo {author} {\bibfnamefont {B.}~\bibnamefont
  {Zhao}}, \bibinfo {author} {\bibfnamefont {J.}~\bibnamefont {Yan}}, \bibinfo
  {author} {\bibfnamefont {D.~G.}\ \bibnamefont {Mandrus}}, \bibinfo {author}
  {\bibfnamefont {D.}~\bibnamefont {Xiao}},\ and\ \bibinfo {author}
  {\bibnamefont {et~al.}},\ }\bibfield  {title} {\bibinfo {title} {Optical
  generation of excitonic valley coherence in monolayer {WS}e$_2$},\ }\href
  {https://doi.org/10.1038/nnano.2013.151} {\bibfield  {journal} {\bibinfo
  {journal} {Nat. Nanotechnol.}\ }\textbf {\bibinfo {volume} {8}},\ \bibinfo
  {pages} {634–638} (\bibinfo {year} {2013})}\BibitemShut {NoStop}%
\bibitem [{\citenamefont {Wang}\ \emph {et~al.}(2016)\citenamefont {Wang},
  \citenamefont {Marie}, \citenamefont {Liu}, \citenamefont {Amand},
  \citenamefont {Robert}, \citenamefont {Cadiz}, \citenamefont {Renucci},\ and\
  \citenamefont {Urbaszek}}]{PhysRevLett.117.187401}%
  \BibitemOpen
  \bibfield  {author} {\bibinfo {author} {\bibfnamefont {G.}~\bibnamefont
  {Wang}}, \bibinfo {author} {\bibfnamefont {X.}~\bibnamefont {Marie}},
  \bibinfo {author} {\bibfnamefont {B.~L.}\ \bibnamefont {Liu}}, \bibinfo
  {author} {\bibfnamefont {T.}~\bibnamefont {Amand}}, \bibinfo {author}
  {\bibfnamefont {C.}~\bibnamefont {Robert}}, \bibinfo {author} {\bibfnamefont
  {F.}~\bibnamefont {Cadiz}}, \bibinfo {author} {\bibfnamefont
  {P.}~\bibnamefont {Renucci}},\ and\ \bibinfo {author} {\bibfnamefont
  {B.}~\bibnamefont {Urbaszek}},\ }\bibfield  {title} {\bibinfo {title}
  {Control of exciton valley coherence in transition metal dichalcogenide
  monolayers},\ }\href {https://doi.org/10.1103/PhysRevLett.117.187401}
  {\bibfield  {journal} {\bibinfo  {journal} {Phys. Rev. Lett.}\ }\textbf
  {\bibinfo {volume} {117}},\ \bibinfo {pages} {187401} (\bibinfo {year}
  {2016})}\BibitemShut {NoStop}%
\bibitem [{\citenamefont {Qiu}\ \emph {et~al.}(2019)\citenamefont {Qiu},
  \citenamefont {Chakraborty}, \citenamefont {Dhara},\ and\ \citenamefont
  {Vamivakas}}]{qiu2019room}%
  \BibitemOpen
  \bibfield  {author} {\bibinfo {author} {\bibfnamefont {L.}~\bibnamefont
  {Qiu}}, \bibinfo {author} {\bibfnamefont {C.}~\bibnamefont {Chakraborty}},
  \bibinfo {author} {\bibfnamefont {S.}~\bibnamefont {Dhara}},\ and\ \bibinfo
  {author} {\bibfnamefont {A.~N.}\ \bibnamefont {Vamivakas}},\ }\bibfield
  {title} {\bibinfo {title} {Room-temperature valley coherence in a polaritonic
  system},\ }\href {https://doi.org/10.1038/s41467-019-09490-6} {\bibfield
  {journal} {\bibinfo  {journal} {Nat. Commun.}\ }\textbf {\bibinfo {volume}
  {10}},\ \bibinfo {pages} {1513} (\bibinfo {year} {2019})}\BibitemShut
  {NoStop}%
\bibitem [{\citenamefont {Hao}\ \emph {et~al.}(2016)\citenamefont {Hao},
  \citenamefont {Moody}, \citenamefont {Wu}, \citenamefont {Dass},
  \citenamefont {Xu}, \citenamefont {Chen}, \citenamefont {Sun}, \citenamefont
  {Li}, \citenamefont {Li}, \citenamefont {Macdonald},\ and\ \citenamefont
  {et~al.}}]{hao_moody2016}%
  \BibitemOpen
  \bibfield  {author} {\bibinfo {author} {\bibfnamefont {K.}~\bibnamefont
  {Hao}}, \bibinfo {author} {\bibfnamefont {G.}~\bibnamefont {Moody}}, \bibinfo
  {author} {\bibfnamefont {F.}~\bibnamefont {Wu}}, \bibinfo {author}
  {\bibfnamefont {C.~K.}\ \bibnamefont {Dass}}, \bibinfo {author}
  {\bibfnamefont {L.}~\bibnamefont {Xu}}, \bibinfo {author} {\bibfnamefont
  {C.-H.}\ \bibnamefont {Chen}}, \bibinfo {author} {\bibfnamefont
  {L.}~\bibnamefont {Sun}}, \bibinfo {author} {\bibfnamefont {M.-Y.}\
  \bibnamefont {Li}}, \bibinfo {author} {\bibfnamefont {L.-J.}\ \bibnamefont
  {Li}}, \bibinfo {author} {\bibfnamefont {A.~H.}\ \bibnamefont {Macdonald}},\
  and\ \bibinfo {author} {\bibnamefont {et~al.}},\ }\bibfield  {title}
  {\bibinfo {title} {Direct measurement of exciton valley coherence in
  monolayer {WSe}$_{2}$},\ }\href {https://doi.org/10.1038/nphys3674}
  {\bibfield  {journal} {\bibinfo  {journal} {Nat. Phys.}\ }\textbf {\bibinfo
  {volume} {12}},\ \bibinfo {pages} {677–682} (\bibinfo {year}
  {2016})}\BibitemShut {NoStop}%
\bibitem [{\citenamefont {Moody}\ \emph {et~al.}(2015)\citenamefont {Moody},
  \citenamefont {Kavir~Dass}, \citenamefont {Hao}, \citenamefont {Chen},
  \citenamefont {Li}, \citenamefont {Singh}, \citenamefont {Tran},
  \citenamefont {Clark}, \citenamefont {Xu}, \citenamefont {Bergh{\"a}user},
  \citenamefont {Malic}, \citenamefont {Knorr},\ and\ \citenamefont
  {Li}}]{moody2015intrinsic}%
  \BibitemOpen
  \bibfield  {author} {\bibinfo {author} {\bibfnamefont {G.}~\bibnamefont
  {Moody}}, \bibinfo {author} {\bibfnamefont {C.}~\bibnamefont {Kavir~Dass}},
  \bibinfo {author} {\bibfnamefont {K.}~\bibnamefont {Hao}}, \bibinfo {author}
  {\bibfnamefont {C.-H.}\ \bibnamefont {Chen}}, \bibinfo {author}
  {\bibfnamefont {L.-J.}\ \bibnamefont {Li}}, \bibinfo {author} {\bibfnamefont
  {A.}~\bibnamefont {Singh}}, \bibinfo {author} {\bibfnamefont
  {K.}~\bibnamefont {Tran}}, \bibinfo {author} {\bibfnamefont {G.}~\bibnamefont
  {Clark}}, \bibinfo {author} {\bibfnamefont {X.}~\bibnamefont {Xu}}, \bibinfo
  {author} {\bibfnamefont {G.}~\bibnamefont {Bergh{\"a}user}}, \bibinfo
  {author} {\bibfnamefont {E.}~\bibnamefont {Malic}}, \bibinfo {author}
  {\bibfnamefont {A.}~\bibnamefont {Knorr}},\ and\ \bibinfo {author}
  {\bibfnamefont {X.}~\bibnamefont {Li}},\ }\bibfield  {title} {\bibinfo
  {title} {Intrinsic homogeneous linewidth and broadening mechanisms of
  excitons in monolayer transition metal dichalcogenides},\ }\href
  {https://doi.org/10.1038/ncomms9315} {\bibfield  {journal} {\bibinfo
  {journal} {Nat. Commun.}\ }\textbf {\bibinfo {volume} {6}},\ \bibinfo {pages}
  {8315} (\bibinfo {year} {2015})}\BibitemShut {NoStop}%
\bibitem [{\citenamefont {Gupta}\ \emph {et~al.}(2023)\citenamefont {Gupta},
  \citenamefont {Watanabe}, \citenamefont {Taniguchi},\ and\ \citenamefont
  {Majumdar}}]{gupta2023}%
  \BibitemOpen
  \bibfield  {author} {\bibinfo {author} {\bibfnamefont {G.}~\bibnamefont
  {Gupta}}, \bibinfo {author} {\bibfnamefont {K.}~\bibnamefont {Watanabe}},
  \bibinfo {author} {\bibfnamefont {T.}~\bibnamefont {Taniguchi}},\ and\
  \bibinfo {author} {\bibfnamefont {K.}~\bibnamefont {Majumdar}},\ }\bibfield
  {title} {\bibinfo {title} {Observation of $~100\%$ valley-coherent excitons
  in monolayer {MoS}$_2$ through giant enhancement of valley coherence time},\
  }\href {https://doi.org/10.1038/s41377-023-01220-4} {\bibfield  {journal}
  {\bibinfo  {journal} {Light-Sci. Appl.}\ }\textbf {\bibinfo {volume} {12}},\
  \bibinfo {pages} {173} (\bibinfo {year} {2023})}\BibitemShut {NoStop}%
\bibitem [{\citenamefont {Lan}\ \emph {et~al.}(2024)\citenamefont {Lan},
  \citenamefont {Xie},\ and\ \citenamefont {Fu}}]{PhysRevB.110.125420}%
  \BibitemOpen
  \bibfield  {author} {\bibinfo {author} {\bibfnamefont {K.}~\bibnamefont
  {Lan}}, \bibinfo {author} {\bibfnamefont {S.}~\bibnamefont {Xie}},\ and\
  \bibinfo {author} {\bibfnamefont {J.}~\bibnamefont {Fu}},\ }\bibfield
  {title} {\bibinfo {title} {Laser-field detuning assisted optimization of
  valley dynamics in monolayer ${\mathrm{wse}}_{2}$},\ }\href
  {https://doi.org/10.1103/PhysRevB.110.125420} {\bibfield  {journal} {\bibinfo
   {journal} {Phys. Rev. B}\ }\textbf {\bibinfo {volume} {110}},\ \bibinfo
  {pages} {125420} (\bibinfo {year} {2024})}\BibitemShut {NoStop}%
\bibitem [{\citenamefont {Wang}\ \emph {et~al.}(2025)\citenamefont {Wang},
  \citenamefont {Shinokita}, \citenamefont {Watanabe}, \citenamefont
  {Taniguchi}, \citenamefont {Konabe},\ and\ \citenamefont
  {Matsuda}}]{acsnano5c02659}%
  \BibitemOpen
  \bibfield  {author} {\bibinfo {author} {\bibfnamefont {H.}~\bibnamefont
  {Wang}}, \bibinfo {author} {\bibfnamefont {K.}~\bibnamefont {Shinokita}},
  \bibinfo {author} {\bibfnamefont {K.}~\bibnamefont {Watanabe}}, \bibinfo
  {author} {\bibfnamefont {T.}~\bibnamefont {Taniguchi}}, \bibinfo {author}
  {\bibfnamefont {S.}~\bibnamefont {Konabe}},\ and\ \bibinfo {author}
  {\bibfnamefont {K.}~\bibnamefont {Matsuda}},\ }\bibfield  {title} {\bibinfo
  {title} {Direct identification of valley coherence and its manipulation in
  monolayer two-dimensional semiconductor},\ }\href
  {https://doi.org/10.1021/acsnano.5c02659} {\bibfield  {journal} {\bibinfo
  {journal} {ACS Nano}\ }\textbf {\bibinfo {volume} {19}},\ \bibinfo {pages}
  {21484} (\bibinfo {year} {2025})}\BibitemShut {NoStop}%
\bibitem [{\citenamefont {Molas}\ \emph {et~al.}(2019)\citenamefont {Molas},
  \citenamefont {Slobodeniuk}, \citenamefont {Kazimierczuk}, \citenamefont
  {Nogajewski}, \citenamefont {Bartos}, \citenamefont {Kapu\ifmmode
  \acute{s}\else \'{s}\fi{}ci\ifmmode~\acute{n}\else \'{n}\fi{}ski},
  \citenamefont {Oreszczuk}, \citenamefont {Watanabe}, \citenamefont
  {Taniguchi}, \citenamefont {Faugeras}, \citenamefont {Kossacki},
  \citenamefont {Basko},\ and\ \citenamefont
  {Potemski}}]{PhysRevLett.123.096803}%
  \BibitemOpen
  \bibfield  {author} {\bibinfo {author} {\bibfnamefont {M.~R.}\ \bibnamefont
  {Molas}}, \bibinfo {author} {\bibfnamefont {A.~O.}\ \bibnamefont
  {Slobodeniuk}}, \bibinfo {author} {\bibfnamefont {T.}~\bibnamefont
  {Kazimierczuk}}, \bibinfo {author} {\bibfnamefont {K.}~\bibnamefont
  {Nogajewski}}, \bibinfo {author} {\bibfnamefont {M.}~\bibnamefont {Bartos}},
  \bibinfo {author} {\bibfnamefont {P.}~\bibnamefont {Kapu\ifmmode
  \acute{s}\else \'{s}\fi{}ci\ifmmode~\acute{n}\else \'{n}\fi{}ski}}, \bibinfo
  {author} {\bibfnamefont {K.}~\bibnamefont {Oreszczuk}}, \bibinfo {author}
  {\bibfnamefont {K.}~\bibnamefont {Watanabe}}, \bibinfo {author}
  {\bibfnamefont {T.}~\bibnamefont {Taniguchi}}, \bibinfo {author}
  {\bibfnamefont {C.}~\bibnamefont {Faugeras}}, \bibinfo {author}
  {\bibfnamefont {P.}~\bibnamefont {Kossacki}}, \bibinfo {author}
  {\bibfnamefont {D.~M.}\ \bibnamefont {Basko}},\ and\ \bibinfo {author}
  {\bibfnamefont {M.}~\bibnamefont {Potemski}},\ }\bibfield  {title} {\bibinfo
  {title} {Probing and manipulating valley coherence of dark excitons in
  monolayer {WSe}$_{2}$},\ }\href
  {https://doi.org/10.1103/PhysRevLett.123.096803} {\bibfield  {journal}
  {\bibinfo  {journal} {Phys. Rev. Lett.}\ }\textbf {\bibinfo {volume} {123}},\
  \bibinfo {pages} {096803} (\bibinfo {year} {2019})}\BibitemShut {NoStop}%
\bibitem [{\citenamefont {Robert}\ \emph {et~al.}(2017)\citenamefont {Robert},
  \citenamefont {Amand}, \citenamefont {Cadiz}, \citenamefont {Lagarde},
  \citenamefont {Courtade}, \citenamefont {Manca}, \citenamefont {Taniguchi},
  \citenamefont {Watanabe}, \citenamefont {Urbaszek},\ and\ \citenamefont
  {Marie}}]{PhysRevB.96.155423}%
  \BibitemOpen
  \bibfield  {author} {\bibinfo {author} {\bibfnamefont {C.}~\bibnamefont
  {Robert}}, \bibinfo {author} {\bibfnamefont {T.}~\bibnamefont {Amand}},
  \bibinfo {author} {\bibfnamefont {F.}~\bibnamefont {Cadiz}}, \bibinfo
  {author} {\bibfnamefont {D.}~\bibnamefont {Lagarde}}, \bibinfo {author}
  {\bibfnamefont {E.}~\bibnamefont {Courtade}}, \bibinfo {author}
  {\bibfnamefont {M.}~\bibnamefont {Manca}}, \bibinfo {author} {\bibfnamefont
  {T.}~\bibnamefont {Taniguchi}}, \bibinfo {author} {\bibfnamefont
  {K.}~\bibnamefont {Watanabe}}, \bibinfo {author} {\bibfnamefont
  {B.}~\bibnamefont {Urbaszek}},\ and\ \bibinfo {author} {\bibfnamefont
  {X.}~\bibnamefont {Marie}},\ }\bibfield  {title} {\bibinfo {title} {Fine
  structure and lifetime of dark excitons in transition metal dichalcogenide
  monolayers},\ }\href {https://doi.org/10.1103/PhysRevB.96.155423} {\bibfield
  {journal} {\bibinfo  {journal} {Phys. Rev. B}\ }\textbf {\bibinfo {volume}
  {96}},\ \bibinfo {pages} {155423} (\bibinfo {year} {2017})}\BibitemShut
  {NoStop}%
\bibitem [{\citenamefont {Ren}\ \emph {et~al.}(2023)\citenamefont {Ren},
  \citenamefont {Robert}, \citenamefont {Glazov}, \citenamefont {Semina},
  \citenamefont {Amand}, \citenamefont {Lombez}, \citenamefont {Lagarde},
  \citenamefont {Taniguchi}, \citenamefont {Watanabe},\ and\ \citenamefont
  {Marie}}]{PhysRevLett.131.116901}%
  \BibitemOpen
  \bibfield  {author} {\bibinfo {author} {\bibfnamefont {L.}~\bibnamefont
  {Ren}}, \bibinfo {author} {\bibfnamefont {C.}~\bibnamefont {Robert}},
  \bibinfo {author} {\bibfnamefont {M.}~\bibnamefont {Glazov}}, \bibinfo
  {author} {\bibfnamefont {M.}~\bibnamefont {Semina}}, \bibinfo {author}
  {\bibfnamefont {T.}~\bibnamefont {Amand}}, \bibinfo {author} {\bibfnamefont
  {L.}~\bibnamefont {Lombez}}, \bibinfo {author} {\bibfnamefont
  {D.}~\bibnamefont {Lagarde}}, \bibinfo {author} {\bibfnamefont
  {T.}~\bibnamefont {Taniguchi}}, \bibinfo {author} {\bibfnamefont
  {K.}~\bibnamefont {Watanabe}},\ and\ \bibinfo {author} {\bibfnamefont
  {X.}~\bibnamefont {Marie}},\ }\bibfield  {title} {\bibinfo {title} {Control
  of the bright-dark exciton splitting using the lamb shift in a
  two-dimensional semiconductor},\ }\href
  {https://doi.org/10.1103/PhysRevLett.131.116901} {\bibfield  {journal}
  {\bibinfo  {journal} {Phys. Rev. Lett.}\ }\textbf {\bibinfo {volume} {131}},\
  \bibinfo {pages} {116901} (\bibinfo {year} {2023})}\BibitemShut {NoStop}%
\bibitem [{\citenamefont {Mad$\acute{\rm{e}}$o}\ \emph
  {et~al.}(2020)\citenamefont {Mad$\acute{\rm{e}}$o}, \citenamefont {Man},
  \citenamefont {Sahoo}, \citenamefont {Campbell}, \citenamefont {Pareek},
  \citenamefont {Wong}, \citenamefont {Al-Mahboob}, \citenamefont {Chan},
  \citenamefont {Karmakar}, \citenamefont {Mariserla}, \citenamefont {Li},
  \citenamefont {Heinz}, \citenamefont {Cao},\ and\ \citenamefont
  {Dani}}]{scienceaba1029}%
  \BibitemOpen
  \bibfield  {author} {\bibinfo {author} {\bibfnamefont {J.}~\bibnamefont
  {Mad$\acute{\rm{e}}$o}}, \bibinfo {author} {\bibfnamefont {M.~K.~L.}\
  \bibnamefont {Man}}, \bibinfo {author} {\bibfnamefont {C.}~\bibnamefont
  {Sahoo}}, \bibinfo {author} {\bibfnamefont {M.}~\bibnamefont {Campbell}},
  \bibinfo {author} {\bibfnamefont {V.}~\bibnamefont {Pareek}}, \bibinfo
  {author} {\bibfnamefont {E.~L.}\ \bibnamefont {Wong}}, \bibinfo {author}
  {\bibfnamefont {A.}~\bibnamefont {Al-Mahboob}}, \bibinfo {author}
  {\bibfnamefont {N.~S.}\ \bibnamefont {Chan}}, \bibinfo {author}
  {\bibfnamefont {A.}~\bibnamefont {Karmakar}}, \bibinfo {author}
  {\bibfnamefont {B.~M.~K.}\ \bibnamefont {Mariserla}}, \bibinfo {author}
  {\bibfnamefont {X.}~\bibnamefont {Li}}, \bibinfo {author} {\bibfnamefont
  {T.~F.}\ \bibnamefont {Heinz}}, \bibinfo {author} {\bibfnamefont
  {T.}~\bibnamefont {Cao}},\ and\ \bibinfo {author} {\bibfnamefont {K.~M.}\
  \bibnamefont {Dani}},\ }\bibfield  {title} {\bibinfo {title} {Directly
  visualizing the momentum-forbidden dark excitons and their dynamics in
  atomically thin semiconductors},\ }\href
  {https://doi.org/10.1126/science.aba1029} {\bibfield  {journal} {\bibinfo
  {journal} {Science}\ }\textbf {\bibinfo {volume} {370}},\ \bibinfo {pages}
  {1199} (\bibinfo {year} {2020})}\BibitemShut {NoStop}%
\bibitem [{foo({\natexlab{a}})}]{footenotedark}%
  \BibitemOpen
  \href@noop {} {} ({\natexlab{a}}),\ \bibinfo {note}
  {\textcolor{blue}{Strictly speaking, bare spin-unlike excitons may exhibit
  weak coupling to out-of-plane polarized light, but this does not modify their
  conventional classification as ``dark excitons'' in the context of
  valley-selective excitation and detection, consistent with the widespread
  usage in TMDC
  literature~\cite{PhysRevLett.123.096803,zhang883,PhysRevResearch.1.032007}}}\BibitemShut
  {NoStop}%
\bibitem [{\citenamefont {Blundo}\ \emph {et~al.}(2022)\citenamefont {Blundo},
  \citenamefont {Junior}, \citenamefont {Surrente}, \citenamefont {Pettinari},
  \citenamefont {Prosnikov}, \citenamefont {Olkowska-Pucko}, \citenamefont
  {Zollner}, \citenamefont {Wo\ifmmode~\acute{z}\else \'{z}\fi{}niak},
  \citenamefont {Chaves}, \citenamefont {Kazimierczuk}, \citenamefont {Felici},
  \citenamefont {Babi\ifmmode~\acute{n}\else \'{n}\fi{}ski}, \citenamefont
  {Molas}, \citenamefont {Christianen}, \citenamefont {Fabian},\ and\
  \citenamefont {Polimeni}}]{PhysRevLett.129.067402}%
  \BibitemOpen
  \bibfield  {author} {\bibinfo {author} {\bibfnamefont {E.}~\bibnamefont
  {Blundo}}, \bibinfo {author} {\bibfnamefont {P.~E.~F.}\ \bibnamefont
  {Junior}}, \bibinfo {author} {\bibfnamefont {A.}~\bibnamefont {Surrente}},
  \bibinfo {author} {\bibfnamefont {G.}~\bibnamefont {Pettinari}}, \bibinfo
  {author} {\bibfnamefont {M.~A.}\ \bibnamefont {Prosnikov}}, \bibinfo {author}
  {\bibfnamefont {K.}~\bibnamefont {Olkowska-Pucko}}, \bibinfo {author}
  {\bibfnamefont {K.}~\bibnamefont {Zollner}}, \bibinfo {author} {\bibfnamefont
  {T.}~\bibnamefont {Wo\ifmmode~\acute{z}\else \'{z}\fi{}niak}}, \bibinfo
  {author} {\bibfnamefont {A.}~\bibnamefont {Chaves}}, \bibinfo {author}
  {\bibfnamefont {T.}~\bibnamefont {Kazimierczuk}}, \bibinfo {author}
  {\bibfnamefont {M.}~\bibnamefont {Felici}}, \bibinfo {author} {\bibfnamefont
  {A.}~\bibnamefont {Babi\ifmmode~\acute{n}\else \'{n}\fi{}ski}}, \bibinfo
  {author} {\bibfnamefont {M.~R.}\ \bibnamefont {Molas}}, \bibinfo {author}
  {\bibfnamefont {P.~C.~M.}\ \bibnamefont {Christianen}}, \bibinfo {author}
  {\bibfnamefont {J.}~\bibnamefont {Fabian}},\ and\ \bibinfo {author}
  {\bibfnamefont {A.}~\bibnamefont {Polimeni}},\ }\bibfield  {title} {\bibinfo
  {title} {Strain-induced exciton hybridization in ${\mathrm{ws}}_{2}$
  monolayers unveiled by zeeman-splitting measurements},\ }\href
  {https://doi.org/10.1103/PhysRevLett.129.067402} {\bibfield  {journal}
  {\bibinfo  {journal} {Phys. Rev. Lett.}\ }\textbf {\bibinfo {volume} {129}},\
  \bibinfo {pages} {067402} (\bibinfo {year} {2022})}\BibitemShut {NoStop}%
\bibitem [{\citenamefont {Jiang}\ \emph {et~al.}(2021)\citenamefont {Jiang},
  \citenamefont {Zheng}, \citenamefont {Lan}, \citenamefont {Saidi},
  \citenamefont {Ren},\ and\ \citenamefont {Zhao}}]{sciadvabf3759}%
  \BibitemOpen
  \bibfield  {author} {\bibinfo {author} {\bibfnamefont {X.}~\bibnamefont
  {Jiang}}, \bibinfo {author} {\bibfnamefont {Q.}~\bibnamefont {Zheng}},
  \bibinfo {author} {\bibfnamefont {Z.}~\bibnamefont {Lan}}, \bibinfo {author}
  {\bibfnamefont {W.~A.}\ \bibnamefont {Saidi}}, \bibinfo {author}
  {\bibfnamefont {X.}~\bibnamefont {Ren}},\ and\ \bibinfo {author}
  {\bibfnamefont {J.}~\bibnamefont {Zhao}},\ }\bibfield  {title} {\bibinfo
  {title} {Real-time \emph{GW}-\rm{BSE} investigations on spin-valley exciton
  dynamics in monolayer transition metal dichalcogenide},\ }\href
  {https://doi.org/10.1126/sciadv.abf3759} {\bibfield  {journal} {\bibinfo
  {journal} {Sci. Adv.}\ }\textbf {\bibinfo {volume} {7}},\ \bibinfo {pages}
  {eabf3759} (\bibinfo {year} {2021})}\BibitemShut {NoStop}%
\bibitem [{\citenamefont {Robert}\ \emph {et~al.}(2020)\citenamefont {Robert},
  \citenamefont {Han}, \citenamefont {Kapuscinski}, \citenamefont {Delhomme},
  \citenamefont {Faugeras}, \citenamefont {Amand}, \citenamefont {Molas},
  \citenamefont {Bartos}, \citenamefont {Watanabe}, \citenamefont {Taniguchi},
  \citenamefont {Urbaszek}, \citenamefont {Potemski},\ and\ \citenamefont
  {Marie}}]{robert2020}%
  \BibitemOpen
  \bibfield  {author} {\bibinfo {author} {\bibfnamefont {C.}~\bibnamefont
  {Robert}}, \bibinfo {author} {\bibfnamefont {B.}~\bibnamefont {Han}},
  \bibinfo {author} {\bibfnamefont {P.}~\bibnamefont {Kapuscinski}}, \bibinfo
  {author} {\bibfnamefont {A.}~\bibnamefont {Delhomme}}, \bibinfo {author}
  {\bibfnamefont {C.}~\bibnamefont {Faugeras}}, \bibinfo {author}
  {\bibfnamefont {T.}~\bibnamefont {Amand}}, \bibinfo {author} {\bibfnamefont
  {M.~R.}\ \bibnamefont {Molas}}, \bibinfo {author} {\bibfnamefont
  {M.}~\bibnamefont {Bartos}}, \bibinfo {author} {\bibfnamefont
  {K.}~\bibnamefont {Watanabe}}, \bibinfo {author} {\bibfnamefont
  {T.}~\bibnamefont {Taniguchi}}, \bibinfo {author} {\bibfnamefont
  {B.}~\bibnamefont {Urbaszek}}, \bibinfo {author} {\bibfnamefont
  {M.}~\bibnamefont {Potemski}},\ and\ \bibinfo {author} {\bibfnamefont
  {X.}~\bibnamefont {Marie}},\ }\bibfield  {title} {\bibinfo {title}
  {Measurement of the spin-forbidden dark excitons in ${\mathrm{mos}}_{2}$ and
  ${\mathrm{mose}}_{2}$ monolayers.},\ }\href
  {https://doi.org/10.1038/s41467-020-17608-4} {\bibfield  {journal} {\bibinfo
  {journal} {Nat. Commun.}\ }\textbf {\bibinfo {volume} {11}},\ \bibinfo
  {pages} {4037} (\bibinfo {year} {2020})}\BibitemShut {NoStop}%
\bibitem [{\citenamefont {Wang}\ and\ \citenamefont {Li}(2008)}]{Wang012106}%
  \BibitemOpen
  \bibfield  {author} {\bibinfo {author} {\bibfnamefont {J.-W.}\ \bibnamefont
  {Wang}}\ and\ \bibinfo {author} {\bibfnamefont {S.-S.}\ \bibnamefont {Li}},\
  }\bibfield  {title} {\bibinfo {title} {Excitonic bright-to-dark transition
  induced by spin-orbit coupling},\ }\href {https://doi.org/10.1063/1.2828861}
  {\bibfield  {journal} {\bibinfo  {journal} {Appl. Phys. Lett.}\ }\textbf
  {\bibinfo {volume} {92}},\ \bibinfo {pages} {012106} (\bibinfo {year}
  {2008})}\BibitemShut {NoStop}%
\bibitem [{\citenamefont {Zhang}\ \emph {et~al.}(2015)\citenamefont {Zhang},
  \citenamefont {You}, \citenamefont {Zhao},\ and\ \citenamefont
  {Heinz}}]{PhysRevLett.115.257403}%
  \BibitemOpen
  \bibfield  {author} {\bibinfo {author} {\bibfnamefont {X.-X.}\ \bibnamefont
  {Zhang}}, \bibinfo {author} {\bibfnamefont {Y.}~\bibnamefont {You}}, \bibinfo
  {author} {\bibfnamefont {S.~Y.~F.}\ \bibnamefont {Zhao}},\ and\ \bibinfo
  {author} {\bibfnamefont {T.~F.}\ \bibnamefont {Heinz}},\ }\bibfield  {title}
  {\bibinfo {title} {Experimental evidence for dark excitons in monolayer
  ${\mathrm{wse}}_{2}$},\ }\href
  {https://doi.org/10.1103/PhysRevLett.115.257403} {\bibfield  {journal}
  {\bibinfo  {journal} {Phys. Rev. Lett.}\ }\textbf {\bibinfo {volume} {115}},\
  \bibinfo {pages} {257403} (\bibinfo {year} {2015})}\BibitemShut {NoStop}%
\bibitem [{\citenamefont {Maialle}\ \emph {et~al.}(1993)\citenamefont
  {Maialle}, \citenamefont {de~Andrada~e Silva},\ and\ \citenamefont
  {Sham}}]{PhysRevB.47.15776}%
  \BibitemOpen
  \bibfield  {author} {\bibinfo {author} {\bibfnamefont {M.~Z.}\ \bibnamefont
  {Maialle}}, \bibinfo {author} {\bibfnamefont {E.~A.}\ \bibnamefont
  {de~Andrada~e Silva}},\ and\ \bibinfo {author} {\bibfnamefont {L.~J.}\
  \bibnamefont {Sham}},\ }\bibfield  {title} {\bibinfo {title} {Exciton spin
  dynamics in quantum wells},\ }\href
  {https://doi.org/10.1103/PhysRevB.47.15776} {\bibfield  {journal} {\bibinfo
  {journal} {Phys. Rev. B}\ }\textbf {\bibinfo {volume} {47}},\ \bibinfo
  {pages} {15776} (\bibinfo {year} {1993})}\BibitemShut {NoStop}%
\bibitem [{foo({\natexlab{b}})}]{footnote-exch}%
  \BibitemOpen
  \href@noop {} {} ({\natexlab{b}}),\ \bibinfo {note} {\textcolor{blue}{The
  effect of the intravalley LR-exchange interaction for bright excitons is
  negligible and is thus neglected~\cite{PhysRevB.89.205303}}.}\BibitemShut
  {Stop}%
\bibitem [{\citenamefont {Ma}\ \emph {et~al.}(2021)\citenamefont {Ma},
  \citenamefont {Wang}, \citenamefont {Han}, \citenamefont {Qu},\ and\
  \citenamefont {Fu}}]{PhysRevB.104.195424}%
  \BibitemOpen
  \bibfield  {author} {\bibinfo {author} {\bibfnamefont {Y.}~\bibnamefont
  {Ma}}, \bibinfo {author} {\bibfnamefont {Q.}~\bibnamefont {Wang}}, \bibinfo
  {author} {\bibfnamefont {S.}~\bibnamefont {Han}}, \bibinfo {author}
  {\bibfnamefont {F.}~\bibnamefont {Qu}},\ and\ \bibinfo {author}
  {\bibfnamefont {J.}~\bibnamefont {Fu}},\ }\bibfield  {title} {\bibinfo
  {title} {Fine structure mediated magnetic response of trion valley
  polarization in monolayer {WSe}$_{2}$},\ }\href
  {https://doi.org/10.1103/PhysRevB.104.195424} {\bibfield  {journal} {\bibinfo
   {journal} {Phys. Rev. B}\ }\textbf {\bibinfo {volume} {104}},\ \bibinfo
  {pages} {195424} (\bibinfo {year} {2021})}\BibitemShut {NoStop}%
\bibitem [{\citenamefont {Qu}\ \emph {et~al.}(2019)\citenamefont {Qu},
  \citenamefont {Bragan{\c{c}}a}, \citenamefont {Vasconcelos}, \citenamefont
  {Liu}, \citenamefont {Xie},\ and\ \citenamefont {Zeng}}]{Qu2019}%
  \BibitemOpen
  \bibfield  {author} {\bibinfo {author} {\bibfnamefont {F.}~\bibnamefont
  {Qu}}, \bibinfo {author} {\bibfnamefont {H.}~\bibnamefont {Bragan{\c{c}}a}},
  \bibinfo {author} {\bibfnamefont {R.}~\bibnamefont {Vasconcelos}}, \bibinfo
  {author} {\bibfnamefont {F.}~\bibnamefont {Liu}}, \bibinfo {author}
  {\bibfnamefont {S.-J.}\ \bibnamefont {Xie}},\ and\ \bibinfo {author}
  {\bibfnamefont {H.}~\bibnamefont {Zeng}},\ }\bibfield  {title} {\bibinfo
  {title} {Controlling valley splitting and polarization of dark- and
  bi-excitons in monolayer {WS}$_2$ by a tilted magnetic field},\ }\href
  {https://doi.org/10.1088/2053-1583/ab2cf7} {\bibfield  {journal} {\bibinfo
  {journal} {2D Mater.}\ }\textbf {\bibinfo {volume} {6}},\ \bibinfo {pages}
  {045014} (\bibinfo {year} {2019})}\BibitemShut {NoStop}%
\bibitem [{\citenamefont {Surrente}\ \emph {et~al.}(2018)\citenamefont
  {Surrente}, \citenamefont {Kłopotowski}, \citenamefont {Zhang},
  \citenamefont {Baranowski}, \citenamefont {Mitioglu}, \citenamefont
  {Ballottin}, \citenamefont {Christianen}, \citenamefont {Dumcenco},
  \citenamefont {Kung}, \citenamefont {Maude}, \citenamefont {Kis},\ and\
  \citenamefont {Plochocka}}]{nanolett.8b01484}%
  \BibitemOpen
  \bibfield  {author} {\bibinfo {author} {\bibfnamefont {A.}~\bibnamefont
  {Surrente}}, \bibinfo {author} {\bibfnamefont {u.}~\bibnamefont
  {Kłopotowski}}, \bibinfo {author} {\bibfnamefont {N.}~\bibnamefont {Zhang}},
  \bibinfo {author} {\bibfnamefont {M.}~\bibnamefont {Baranowski}}, \bibinfo
  {author} {\bibfnamefont {A.~A.}\ \bibnamefont {Mitioglu}}, \bibinfo {author}
  {\bibfnamefont {M.~V.}\ \bibnamefont {Ballottin}}, \bibinfo {author}
  {\bibfnamefont {P.~C.}\ \bibnamefont {Christianen}}, \bibinfo {author}
  {\bibfnamefont {D.}~\bibnamefont {Dumcenco}}, \bibinfo {author}
  {\bibfnamefont {Y.-C.}\ \bibnamefont {Kung}}, \bibinfo {author}
  {\bibfnamefont {D.~K.}\ \bibnamefont {Maude}}, \bibinfo {author}
  {\bibfnamefont {A.}~\bibnamefont {Kis}},\ and\ \bibinfo {author}
  {\bibfnamefont {P.}~\bibnamefont {Plochocka}},\ }\bibfield  {title} {\bibinfo
  {title} {Intervalley scattering of interlayer excitons in a
  {MoS}$_2$/{{MoSe}}$_2$/{MoS}$_2$ heterostructure in high magnetic field},\
  }\href {https://doi.org/10.1021/acs.nanolett.8b01484} {\bibfield  {journal}
  {\bibinfo  {journal} {Nano Lett.}\ }\textbf {\bibinfo {volume} {18}},\
  \bibinfo {pages} {3994} (\bibinfo {year} {2018})}\BibitemShut {NoStop}%
\bibitem [{\citenamefont {Palummo}\ \emph {et~al.}(2015)\citenamefont
  {Palummo}, \citenamefont {Bernardi},\ and\ \citenamefont
  {Grossman}}]{nl503799t}%
  \BibitemOpen
  \bibfield  {author} {\bibinfo {author} {\bibfnamefont {M.}~\bibnamefont
  {Palummo}}, \bibinfo {author} {\bibfnamefont {M.}~\bibnamefont {Bernardi}},\
  and\ \bibinfo {author} {\bibfnamefont {J.~C.}\ \bibnamefont {Grossman}},\
  }\bibfield  {title} {\bibinfo {title} {Exciton radiative lifetimes in
  two-dimensional transition metal dichalcogenides},\ }\href
  {https://doi.org/10.1021/nl503799t} {\bibfield  {journal} {\bibinfo
  {journal} {Nano Lett.}\ }\textbf {\bibinfo {volume} {15}},\ \bibinfo {pages}
  {2794} (\bibinfo {year} {2015})}\BibitemShut {NoStop}%
\bibitem [{\citenamefont {Zeng}\ \emph {et~al.}(2012)\citenamefont {Zeng},
  \citenamefont {Dai}, \citenamefont {Yao}, \citenamefont {Xiao},\ and\
  \citenamefont {Cui}}]{zeng2012valley}%
  \BibitemOpen
  \bibfield  {author} {\bibinfo {author} {\bibfnamefont {H.}~\bibnamefont
  {Zeng}}, \bibinfo {author} {\bibfnamefont {J.}~\bibnamefont {Dai}}, \bibinfo
  {author} {\bibfnamefont {W.}~\bibnamefont {Yao}}, \bibinfo {author}
  {\bibfnamefont {D.}~\bibnamefont {Xiao}},\ and\ \bibinfo {author}
  {\bibfnamefont {X.}~\bibnamefont {Cui}},\ }\bibfield  {title} {\bibinfo
  {title} {Valley polarization in {MoS}$_2$ monolayers by optical pumping},\
  }\href {https://doi.org/10.1038/nnano.2012.95} {\bibfield  {journal}
  {\bibinfo  {journal} {Nat. Nanotechnol.}\ }\textbf {\bibinfo {volume} {7}},\
  \bibinfo {pages} {490} (\bibinfo {year} {2012})}\BibitemShut {NoStop}%
\bibitem [{\citenamefont {Fu}\ \emph {et~al.}(2018)\citenamefont {Fu},
  \citenamefont {Bezerra},\ and\ \citenamefont {Qu}}]{PhysRevB.97.115425}%
  \BibitemOpen
  \bibfield  {author} {\bibinfo {author} {\bibfnamefont {J.}~\bibnamefont
  {Fu}}, \bibinfo {author} {\bibfnamefont {A.}~\bibnamefont {Bezerra}},\ and\
  \bibinfo {author} {\bibfnamefont {F.}~\bibnamefont {Qu}},\ }\bibfield
  {title} {\bibinfo {title} {Valley dynamics of intravalley and intervalley
  multiexcitonic states in monolayer {WS}$_{2}$},\ }\href
  {https://doi.org/10.1103/PhysRevB.97.115425} {\bibfield  {journal} {\bibinfo
  {journal} {Phys. Rev. B}\ }\textbf {\bibinfo {volume} {97}},\ \bibinfo
  {pages} {115425} (\bibinfo {year} {2018})}\BibitemShut {NoStop}%
\bibitem [{\citenamefont {Dias}\ \emph {et~al.}(2020)\citenamefont {Dias},
  \citenamefont {Bragan\ifmmode~\mbox{\c{c}}\else \c{c}\fi{}a}, \citenamefont
  {Zeng}, \citenamefont {Fonseca}, \citenamefont {Liu},\ and\ \citenamefont
  {Qu}}]{PhysRevB.101.085406}%
  \BibitemOpen
  \bibfield  {author} {\bibinfo {author} {\bibfnamefont {A.~C.}\ \bibnamefont
  {Dias}}, \bibinfo {author} {\bibfnamefont {H.}~\bibnamefont
  {Bragan\ifmmode~\mbox{\c{c}}\else \c{c}\fi{}a}}, \bibinfo {author}
  {\bibfnamefont {H.}~\bibnamefont {Zeng}}, \bibinfo {author} {\bibfnamefont
  {A.~L.~A.}\ \bibnamefont {Fonseca}}, \bibinfo {author} {\bibfnamefont
  {D.-S.}\ \bibnamefont {Liu}},\ and\ \bibinfo {author} {\bibfnamefont
  {F.}~\bibnamefont {Qu}},\ }\bibfield  {title} {\bibinfo {title} {Large
  room-temperature valley polarization by valley-selective switching of exciton
  ground state},\ }\href {https://doi.org/10.1103/PhysRevB.101.085406}
  {\bibfield  {journal} {\bibinfo  {journal} {Phys. Rev. B}\ }\textbf {\bibinfo
  {volume} {101}},\ \bibinfo {pages} {085406} (\bibinfo {year}
  {2020})}\BibitemShut {NoStop}%
\bibitem [{\citenamefont {Fu}\ \emph {et~al.}(2019)\citenamefont {Fu},
  \citenamefont {Cruz},\ and\ \citenamefont {Qu}}]{10.1063/1.5112823}%
  \BibitemOpen
  \bibfield  {author} {\bibinfo {author} {\bibfnamefont {J.}~\bibnamefont
  {Fu}}, \bibinfo {author} {\bibfnamefont {J.~M.~R.}\ \bibnamefont {Cruz}},\
  and\ \bibinfo {author} {\bibfnamefont {F.}~\bibnamefont {Qu}},\ }\bibfield
  {title} {\bibinfo {title} {Valley dynamics of different trion species in
  monolayer wse2},\ }\href {https://doi.org/10.1063/1.5112823} {\bibfield
  {journal} {\bibinfo  {journal} {Appl. Phys. Lett.}\ }\textbf {\bibinfo
  {volume} {115}},\ \bibinfo {pages} {082101} (\bibinfo {year}
  {2019})}\BibitemShut {NoStop}%
\bibitem [{\citenamefont {Baranowski}\ \emph {et~al.}(2017)\citenamefont
  {Baranowski}, \citenamefont {Surrente}, \citenamefont {Maude}, \citenamefont
  {Ballottin}, \citenamefont {Mitioglu}, \citenamefont {Christianen},
  \citenamefont {Kung}, \citenamefont {Dumcenco}, \citenamefont {Kis},\ and\
  \citenamefont {Plochocka}}]{Baranowski2017}%
  \BibitemOpen
  \bibfield  {author} {\bibinfo {author} {\bibfnamefont {M.}~\bibnamefont
  {Baranowski}}, \bibinfo {author} {\bibfnamefont {A.}~\bibnamefont
  {Surrente}}, \bibinfo {author} {\bibfnamefont {D.~K.}\ \bibnamefont {Maude}},
  \bibinfo {author} {\bibfnamefont {M.}~\bibnamefont {Ballottin}}, \bibinfo
  {author} {\bibfnamefont {A.~A.}\ \bibnamefont {Mitioglu}}, \bibinfo {author}
  {\bibfnamefont {P.~C.~M.}\ \bibnamefont {Christianen}}, \bibinfo {author}
  {\bibfnamefont {Y.~C.}\ \bibnamefont {Kung}}, \bibinfo {author}
  {\bibfnamefont {D.}~\bibnamefont {Dumcenco}}, \bibinfo {author}
  {\bibfnamefont {A.}~\bibnamefont {Kis}},\ and\ \bibinfo {author}
  {\bibfnamefont {P.}~\bibnamefont {Plochocka}},\ }\bibfield  {title} {\bibinfo
  {title} {Dark excitons and the elusive valley polarization in transition
  metal dichalcogenides},\ }\href {https://doi.org/10.1088/2053-1583/aa58a0}
  {\bibfield  {journal} {\bibinfo  {journal} {2D Mater.}\ }\textbf {\bibinfo
  {volume} {4}},\ \bibinfo {pages} {025016} (\bibinfo {year}
  {2017})}\BibitemShut {NoStop}%
\bibitem [{foo({\natexlab{c}})}]{footnote-impurity}%
  \BibitemOpen
  \href@noop {} {} ({\natexlab{c}}),\ \bibinfo {note}
  {\textcolor{blue}{Momentum-randomizing scattering (e.g., due to impurities)
  is encoded into momentum relaxation time (see the SM), which, in the presence
  of the $\mathbf{k}$-dependent LR exchange field, yields the
  Maialle-Silva-Sham intervalley decoherence rate~\cite{PhysRevB.47.15776},
  analogous to D'yankonov-Perel' spin relaxation~\cite{1986JETP63655D}.
  Impurity degrees of freedom are thus omitted from the
  Hamiltonian}.}\BibitemShut {Stop}%
\bibitem [{\citenamefont {Baumgratz}\ \emph {et~al.}(2014)\citenamefont
  {Baumgratz}, \citenamefont {Cramer},\ and\ \citenamefont
  {Plenio}}]{PhysRevLett.113.140401}%
  \BibitemOpen
  \bibfield  {author} {\bibinfo {author} {\bibfnamefont {T.}~\bibnamefont
  {Baumgratz}}, \bibinfo {author} {\bibfnamefont {M.}~\bibnamefont {Cramer}},\
  and\ \bibinfo {author} {\bibfnamefont {M.~B.}\ \bibnamefont {Plenio}},\
  }\bibfield  {title} {\bibinfo {title} {Quantifying coherence},\ }\href
  {https://doi.org/10.1103/PhysRevLett.113.140401} {\bibfield  {journal}
  {\bibinfo  {journal} {Phys. Rev. Lett.}\ }\textbf {\bibinfo {volume} {113}},\
  \bibinfo {pages} {140401} (\bibinfo {year} {2014})}\BibitemShut {NoStop}%
\bibitem [{\citenamefont {Jha}\ \emph {et~al.}(2018)\citenamefont {Jha},
  \citenamefont {Shitrit}, \citenamefont {Ren}, \citenamefont {Wang},\ and\
  \citenamefont {Zhang}}]{PhysRevLett.121.116102}%
  \BibitemOpen
  \bibfield  {author} {\bibinfo {author} {\bibfnamefont {P.~K.}\ \bibnamefont
  {Jha}}, \bibinfo {author} {\bibfnamefont {N.}~\bibnamefont {Shitrit}},
  \bibinfo {author} {\bibfnamefont {X.}~\bibnamefont {Ren}}, \bibinfo {author}
  {\bibfnamefont {Y.}~\bibnamefont {Wang}},\ and\ \bibinfo {author}
  {\bibfnamefont {X.}~\bibnamefont {Zhang}},\ }\bibfield  {title} {\bibinfo
  {title} {Spontaneous exciton valley coherence in transition metal
  dichalcogenide monolayers interfaced with an anisotropic metasurface},\
  }\href {https://doi.org/10.1103/PhysRevLett.121.116102} {\bibfield  {journal}
  {\bibinfo  {journal} {Phys. Rev. Lett.}\ }\textbf {\bibinfo {volume} {121}},\
  \bibinfo {pages} {116102} (\bibinfo {year} {2018})}\BibitemShut {NoStop}%
\bibitem [{foo({\natexlab{d}})}]{footenotetime}%
  \BibitemOpen
  \href@noop {} {} ({\natexlab{d}}),\ \bibinfo {note} {the coherence time is
  also widely defined as a time over which the coherence intensity decays to
  $e^{-1}$ of its initial value.}\BibitemShut {Stop}%
\bibitem [{\citenamefont {Chen}\ \emph {et~al.}(2020)\citenamefont {Chen},
  \citenamefont {Zhang}, \citenamefont {Ma}, \citenamefont {Zhang},
  \citenamefont {Li}, \citenamefont {Zhang}, \citenamefont {Zeng},
  \citenamefont {Zhan}, \citenamefont {He}, \citenamefont {Ren}, \citenamefont
  {Cheng},\ and\ \citenamefont {Liu}}]{Chen20}%
  \BibitemOpen
  \bibfield  {author} {\bibinfo {author} {\bibfnamefont {C.}~\bibnamefont
  {Chen}}, \bibinfo {author} {\bibfnamefont {Y.}~\bibnamefont {Zhang}},
  \bibinfo {author} {\bibfnamefont {L.}~\bibnamefont {Ma}}, \bibinfo {author}
  {\bibfnamefont {Y.}~\bibnamefont {Zhang}}, \bibinfo {author} {\bibfnamefont
  {Z.}~\bibnamefont {Li}}, \bibinfo {author} {\bibfnamefont {R.}~\bibnamefont
  {Zhang}}, \bibinfo {author} {\bibfnamefont {X.}~\bibnamefont {Zeng}},
  \bibinfo {author} {\bibfnamefont {Z.}~\bibnamefont {Zhan}}, \bibinfo {author}
  {\bibfnamefont {C.}~\bibnamefont {He}}, \bibinfo {author} {\bibfnamefont
  {X.}~\bibnamefont {Ren}}, \bibinfo {author} {\bibfnamefont {C.}~\bibnamefont
  {Cheng}},\ and\ \bibinfo {author} {\bibfnamefont {C.}~\bibnamefont {Liu}},\
  }\bibfield  {title} {\bibinfo {title} {Flexible generation of higher-order
  poincar$\acute{\rm{e}}$ beams with high efficiency by manipulating the two
  eigenstates of polarized optical vortices},\ }\href
  {https://doi.org/10.1364/OE.388727} {\bibfield  {journal} {\bibinfo
  {journal} {Opt. Express}\ }\textbf {\bibinfo {volume} {28}},\ \bibinfo
  {pages} {10618} (\bibinfo {year} {2020})}\BibitemShut {NoStop}%
\bibitem [{\citenamefont {Robert}\ \emph {et~al.}(2021)\citenamefont {Robert},
  \citenamefont {Dery}, \citenamefont {Ren}, \citenamefont {Van~Tuan},
  \citenamefont {Courtade}, \citenamefont {Yang}, \citenamefont {Urbaszek},
  \citenamefont {Lagarde}, \citenamefont {Watanabe}, \citenamefont {Taniguchi},
  \citenamefont {Amand},\ and\ \citenamefont {Marie}}]{PhysRevLett.126.067403}%
  \BibitemOpen
  \bibfield  {author} {\bibinfo {author} {\bibfnamefont {C.}~\bibnamefont
  {Robert}}, \bibinfo {author} {\bibfnamefont {H.}~\bibnamefont {Dery}},
  \bibinfo {author} {\bibfnamefont {L.}~\bibnamefont {Ren}}, \bibinfo {author}
  {\bibfnamefont {D.}~\bibnamefont {Van~Tuan}}, \bibinfo {author}
  {\bibfnamefont {E.}~\bibnamefont {Courtade}}, \bibinfo {author}
  {\bibfnamefont {M.}~\bibnamefont {Yang}}, \bibinfo {author} {\bibfnamefont
  {B.}~\bibnamefont {Urbaszek}}, \bibinfo {author} {\bibfnamefont
  {D.}~\bibnamefont {Lagarde}}, \bibinfo {author} {\bibfnamefont
  {K.}~\bibnamefont {Watanabe}}, \bibinfo {author} {\bibfnamefont
  {T.}~\bibnamefont {Taniguchi}}, \bibinfo {author} {\bibfnamefont
  {T.}~\bibnamefont {Amand}},\ and\ \bibinfo {author} {\bibfnamefont
  {X.}~\bibnamefont {Marie}},\ }\bibfield  {title} {\bibinfo {title}
  {Measurement of conduction and valence bands $g$-factors in a transition
  metal dichalcogenide monolayer},\ }\href
  {https://doi.org/10.1103/PhysRevLett.126.067403} {\bibfield  {journal}
  {\bibinfo  {journal} {Phys. Rev. Lett.}\ }\textbf {\bibinfo {volume} {126}},\
  \bibinfo {pages} {067403} (\bibinfo {year} {2021})}\BibitemShut {NoStop}%
\bibitem [{\citenamefont {Aivazian}\ \emph {et~al.}(2015)\citenamefont
  {Aivazian}, \citenamefont {Gong}, \citenamefont {Jones}, \citenamefont {Chu},
  \citenamefont {Yan}, \citenamefont {Mandrus}, \citenamefont {Zhang},
  \citenamefont {Cobden}, \citenamefont {Yao},\ and\ \citenamefont
  {Xu}}]{aivazianmagnetic2015}%
  \BibitemOpen
  \bibfield  {author} {\bibinfo {author} {\bibfnamefont {G.}~\bibnamefont
  {Aivazian}}, \bibinfo {author} {\bibfnamefont {Z.}~\bibnamefont {Gong}},
  \bibinfo {author} {\bibfnamefont {A.~M.}\ \bibnamefont {Jones}}, \bibinfo
  {author} {\bibfnamefont {R.-L.}\ \bibnamefont {Chu}}, \bibinfo {author}
  {\bibfnamefont {J.}~\bibnamefont {Yan}}, \bibinfo {author} {\bibfnamefont
  {D.~G.}\ \bibnamefont {Mandrus}}, \bibinfo {author} {\bibfnamefont
  {C.}~\bibnamefont {Zhang}}, \bibinfo {author} {\bibfnamefont
  {D.}~\bibnamefont {Cobden}}, \bibinfo {author} {\bibfnamefont
  {W.}~\bibnamefont {Yao}},\ and\ \bibinfo {author} {\bibfnamefont
  {X.}~\bibnamefont {Xu}},\ }\bibfield  {title} {\bibinfo {title} {Magnetic
  control of valley pseudospin in monolayer {WSe}$_2$},\ }\href
  {https://doi.org/10.1038/nphys3201} {\bibfield  {journal} {\bibinfo
  {journal} {Nat. Phys.}\ }\textbf {\bibinfo {volume} {11}},\ \bibinfo {pages}
  {148} (\bibinfo {year} {2015})}\BibitemShut {NoStop}%
\bibitem [{\citenamefont {Förste}\ \emph {et~al.}(2020)\citenamefont
  {Förste}, \citenamefont {Tepliakov}, \citenamefont {Kruchinin},
  \citenamefont {Lindlau}, \citenamefont {Funk}, \citenamefont {Förg},
  \citenamefont {Watanabe}, \citenamefont {Taniguchi}, \citenamefont
  {Baimuratov},\ and\ \citenamefont {Högele}}]{tepliakov2020}%
  \BibitemOpen
  \bibfield  {author} {\bibinfo {author} {\bibfnamefont {J.}~\bibnamefont
  {Förste}}, \bibinfo {author} {\bibfnamefont {N.~V.}\ \bibnamefont
  {Tepliakov}}, \bibinfo {author} {\bibfnamefont {S.~Y.}\ \bibnamefont
  {Kruchinin}}, \bibinfo {author} {\bibfnamefont {J.}~\bibnamefont {Lindlau}},
  \bibinfo {author} {\bibfnamefont {V.}~\bibnamefont {Funk}}, \bibinfo {author}
  {\bibfnamefont {M.}~\bibnamefont {Förg}}, \bibinfo {author} {\bibfnamefont
  {K.}~\bibnamefont {Watanabe}}, \bibinfo {author} {\bibfnamefont
  {T.}~\bibnamefont {Taniguchi}}, \bibinfo {author} {\bibfnamefont {A.~S.}\
  \bibnamefont {Baimuratov}},\ and\ \bibinfo {author} {\bibfnamefont
  {A.}~\bibnamefont {Högele}},\ }\bibfield  {title} {\bibinfo {title} {Exciton
  g-factors in monolayer and bilayer {WSe}$_2$ from experiment and theory},\
  }\href {https://doi.org/10.1038/s41467-020-18019-1} {\bibfield  {journal}
  {\bibinfo  {journal} {Nat. Commun.}\ }\textbf {\bibinfo {volume} {11}},\
  \bibinfo {pages} {4539} (\bibinfo {year} {2020})}\BibitemShut {NoStop}%
\bibitem [{\citenamefont {Deilmann}\ \emph {et~al.}(2020)\citenamefont
  {Deilmann}, \citenamefont {Kr\"uger},\ and\ \citenamefont
  {Rohlfing}}]{PhysRevLett.124.226402}%
  \BibitemOpen
  \bibfield  {author} {\bibinfo {author} {\bibfnamefont {T.}~\bibnamefont
  {Deilmann}}, \bibinfo {author} {\bibfnamefont {P.}~\bibnamefont {Kr\"uger}},\
  and\ \bibinfo {author} {\bibfnamefont {M.}~\bibnamefont {Rohlfing}},\
  }\bibfield  {title} {\bibinfo {title} {Ab initio studies of exciton $g$
  factors: Monolayer transition metal dichalcogenides in magnetic fields},\
  }\href {https://doi.org/10.1103/PhysRevLett.124.226402} {\bibfield  {journal}
  {\bibinfo  {journal} {Phys. Rev. Lett.}\ }\textbf {\bibinfo {volume} {124}},\
  \bibinfo {pages} {226402} (\bibinfo {year} {2020})}\BibitemShut {NoStop}%
\bibitem [{\citenamefont {Wo\ifmmode~\acute{z}\else \'{z}\fi{}niak}\ \emph
  {et~al.}(2020)\citenamefont {Wo\ifmmode~\acute{z}\else \'{z}\fi{}niak},
  \citenamefont {Faria~Junior}, \citenamefont {Seifert}, \citenamefont
  {Chaves},\ and\ \citenamefont {Kunstmann}}]{PhysRevB.101.235408}%
  \BibitemOpen
  \bibfield  {author} {\bibinfo {author} {\bibfnamefont {T.}~\bibnamefont
  {Wo\ifmmode~\acute{z}\else \'{z}\fi{}niak}}, \bibinfo {author} {\bibfnamefont
  {P.~E.}\ \bibnamefont {Faria~Junior}}, \bibinfo {author} {\bibfnamefont
  {G.}~\bibnamefont {Seifert}}, \bibinfo {author} {\bibfnamefont
  {A.}~\bibnamefont {Chaves}},\ and\ \bibinfo {author} {\bibfnamefont
  {J.}~\bibnamefont {Kunstmann}},\ }\bibfield  {title} {\bibinfo {title}
  {Exciton $g$ factors of van der waals heterostructures from first-principles
  calculations},\ }\href {https://doi.org/10.1103/PhysRevB.101.235408}
  {\bibfield  {journal} {\bibinfo  {journal} {Phys. Rev. B}\ }\textbf {\bibinfo
  {volume} {101}},\ \bibinfo {pages} {235408} (\bibinfo {year}
  {2020})}\BibitemShut {NoStop}%
\bibitem [{\citenamefont {Junior}\ \emph {et~al.}(2022)\citenamefont {Junior},
  \citenamefont {Zollner}, \citenamefont {Woźniak}, \citenamefont {Kurpas},
  \citenamefont {Gmitra},\ and\ \citenamefont {Fabian}}]{Faria2022}%
  \BibitemOpen
  \bibfield  {author} {\bibinfo {author} {\bibfnamefont {P.~E.~F.}\
  \bibnamefont {Junior}}, \bibinfo {author} {\bibfnamefont {K.}~\bibnamefont
  {Zollner}}, \bibinfo {author} {\bibfnamefont {T.}~\bibnamefont {Woźniak}},
  \bibinfo {author} {\bibfnamefont {M.}~\bibnamefont {Kurpas}}, \bibinfo
  {author} {\bibfnamefont {M.}~\bibnamefont {Gmitra}},\ and\ \bibinfo {author}
  {\bibfnamefont {J.}~\bibnamefont {Fabian}},\ }\bibfield  {title} {\bibinfo
  {title} {First-principles insights into the spin-valley physics of strained
  transition metal dichalcogenides monolayers},\ }\href
  {https://doi.org/10.1088/1367-2630/ac7e21} {\bibfield  {journal} {\bibinfo
  {journal} {New J. Phys.}\ }\textbf {\bibinfo {volume} {24}},\ \bibinfo
  {pages} {083004} (\bibinfo {year} {2022})}\BibitemShut {NoStop}%
\bibitem [{\citenamefont {Wang}\ \emph {et~al.}(2014)\citenamefont {Wang},
  \citenamefont {Bouet}, \citenamefont {Lagarde}, \citenamefont {Vidal},
  \citenamefont {Balocchi}, \citenamefont {Amand}, \citenamefont {Marie},\ and\
  \citenamefont {Urbaszek}}]{PhysRevB.90.075413}%
  \BibitemOpen
  \bibfield  {author} {\bibinfo {author} {\bibfnamefont {G.}~\bibnamefont
  {Wang}}, \bibinfo {author} {\bibfnamefont {L.}~\bibnamefont {Bouet}},
  \bibinfo {author} {\bibfnamefont {D.}~\bibnamefont {Lagarde}}, \bibinfo
  {author} {\bibfnamefont {M.}~\bibnamefont {Vidal}}, \bibinfo {author}
  {\bibfnamefont {A.}~\bibnamefont {Balocchi}}, \bibinfo {author}
  {\bibfnamefont {T.}~\bibnamefont {Amand}}, \bibinfo {author} {\bibfnamefont
  {X.}~\bibnamefont {Marie}},\ and\ \bibinfo {author} {\bibfnamefont
  {B.}~\bibnamefont {Urbaszek}},\ }\bibfield  {title} {\bibinfo {title} {Valley
  dynamics probed through charged and neutral exciton emission in monolayer
  ${\mathrm{wse}}_{2}$},\ }\href {https://doi.org/10.1103/PhysRevB.90.075413}
  {\bibfield  {journal} {\bibinfo  {journal} {Phys. Rev. B}\ }\textbf {\bibinfo
  {volume} {90}},\ \bibinfo {pages} {075413} (\bibinfo {year}
  {2014})}\BibitemShut {NoStop}%
\bibitem [{\citenamefont {Zhang}\ \emph {et~al.}(2017)\citenamefont {Zhang},
  \citenamefont {Cao}, \citenamefont {Lu}, \citenamefont {Lin}, \citenamefont
  {Zhang}, \citenamefont {Wang}, \citenamefont {Li}, \citenamefont {Hone},
  \citenamefont {Robinson}, \citenamefont {Smirnov}, \citenamefont {Louie},\
  and\ \citenamefont {Heinz}}]{zhang883}%
  \BibitemOpen
  \bibfield  {author} {\bibinfo {author} {\bibfnamefont {X.-X.}\ \bibnamefont
  {Zhang}}, \bibinfo {author} {\bibfnamefont {T.}~\bibnamefont {Cao}}, \bibinfo
  {author} {\bibfnamefont {Z.}~\bibnamefont {Lu}}, \bibinfo {author}
  {\bibfnamefont {Y.-C.}\ \bibnamefont {Lin}}, \bibinfo {author} {\bibfnamefont
  {F.}~\bibnamefont {Zhang}}, \bibinfo {author} {\bibfnamefont
  {Y.}~\bibnamefont {Wang}}, \bibinfo {author} {\bibfnamefont {Z.}~\bibnamefont
  {Li}}, \bibinfo {author} {\bibfnamefont {J.~C.}\ \bibnamefont {Hone}},
  \bibinfo {author} {\bibfnamefont {J.~A.}\ \bibnamefont {Robinson}}, \bibinfo
  {author} {\bibfnamefont {D.}~\bibnamefont {Smirnov}}, \bibinfo {author}
  {\bibfnamefont {S.~G.}\ \bibnamefont {Louie}},\ and\ \bibinfo {author}
  {\bibfnamefont {T.~F.}\ \bibnamefont {Heinz}},\ }\bibfield  {title} {\bibinfo
  {title} {Magnetic brightening and control of dark excitons in monolayer
  wse2},\ }\href {https://doi.org/10.1038/nnano.2017.105} {\bibfield  {journal}
  {\bibinfo  {journal} {Nat. Nanotechnol.}\ }\textbf {\bibinfo {volume} {12}},\
  \bibinfo {pages} {883} (\bibinfo {year} {2017})}\BibitemShut {NoStop}%
\bibitem [{\citenamefont {Ma}\ \emph {et~al.}(2022)\citenamefont {Ma},
  \citenamefont {Kudtarkar}, \citenamefont {Chen}, \citenamefont {Cunha},
  \citenamefont {Ma}, \citenamefont {Watanabe}, \citenamefont {Taniguchi},
  \citenamefont {Qian}, \citenamefont {Hipwell}, \citenamefont {Wong},\ and\
  \citenamefont {Lan}}]{Xuezhi2022}%
  \BibitemOpen
  \bibfield  {author} {\bibinfo {author} {\bibfnamefont {X.}~\bibnamefont
  {Ma}}, \bibinfo {author} {\bibfnamefont {K.}~\bibnamefont {Kudtarkar}},
  \bibinfo {author} {\bibfnamefont {Y.}~\bibnamefont {Chen}}, \bibinfo {author}
  {\bibfnamefont {P.}~\bibnamefont {Cunha}}, \bibinfo {author} {\bibfnamefont
  {Y.}~\bibnamefont {Ma}}, \bibinfo {author} {\bibfnamefont {K.}~\bibnamefont
  {Watanabe}}, \bibinfo {author} {\bibfnamefont {T.}~\bibnamefont {Taniguchi}},
  \bibinfo {author} {\bibfnamefont {X.}~\bibnamefont {Qian}}, \bibinfo {author}
  {\bibfnamefont {M.~C.}\ \bibnamefont {Hipwell}}, \bibinfo {author}
  {\bibfnamefont {Z.~J.}\ \bibnamefont {Wong}},\ and\ \bibinfo {author}
  {\bibfnamefont {S.}~\bibnamefont {Lan}},\ }\bibfield  {title} {\bibinfo
  {title} {Coherent momentum control of forbidden excitons},\ }\href
  {https://doi.org/10.1038/s41467-022-34740-5} {\bibfield  {journal} {\bibinfo
  {journal} {Nat. Commun.}\ }\textbf {\bibinfo {volume} {13}},\ \bibinfo
  {pages} {6916} (\bibinfo {year} {2022})}\BibitemShut {NoStop}%
\bibitem [{\citenamefont {Tang}\ \emph {et~al.}(2019)\citenamefont {Tang},
  \citenamefont {Mak},\ and\ \citenamefont {Shan}}]{Tang2019}%
  \BibitemOpen
  \bibfield  {author} {\bibinfo {author} {\bibfnamefont {Y.}~\bibnamefont
  {Tang}}, \bibinfo {author} {\bibfnamefont {K.~F.}\ \bibnamefont {Mak}},\ and\
  \bibinfo {author} {\bibfnamefont {J.}~\bibnamefont {Shan}},\ }\bibfield
  {title} {\bibinfo {title} {Long valley lifetime of dark excitons in
  single-layer wse2},\ }\href {https://doi.org/10.1038/s41467-019-12129-1}
  {\bibfield  {journal} {\bibinfo  {journal} {Nat. Commun.}\ }\textbf {\bibinfo
  {volume} {10}},\ \bibinfo {pages} {4047} (\bibinfo {year}
  {2019})}\BibitemShut {NoStop}%
\bibitem [{\citenamefont {Gomez~Sanchez}\ \emph {et~al.}(2024)\citenamefont
  {Gomez~Sanchez}, \citenamefont {Peng}, \citenamefont {Li}, \citenamefont
  {Shih}, \citenamefont {Chien},\ and\ \citenamefont
  {Cheng}}]{acsnano.4c01881}%
  \BibitemOpen
  \bibfield  {author} {\bibinfo {author} {\bibfnamefont {O.~J.}\ \bibnamefont
  {Gomez~Sanchez}}, \bibinfo {author} {\bibfnamefont {G.-H.}\ \bibnamefont
  {Peng}}, \bibinfo {author} {\bibfnamefont {W.-H.}\ \bibnamefont {Li}},
  \bibinfo {author} {\bibfnamefont {C.-H.}\ \bibnamefont {Shih}}, \bibinfo
  {author} {\bibfnamefont {C.-H.}\ \bibnamefont {Chien}},\ and\ \bibinfo
  {author} {\bibfnamefont {S.-J.}\ \bibnamefont {Cheng}},\ }\bibfield  {title}
  {\bibinfo {title} {Enhanced photo-excitation and angular-momentum imprint of
  gray excitons in wse2 monolayers by spin–orbit-coupled vector vortex
  beams},\ }\href {https://doi.org/10.1021/acsnano.4c01881} {\bibfield
  {journal} {\bibinfo  {journal} {ACS Nano}\ }\textbf {\bibinfo {volume}
  {18}},\ \bibinfo {pages} {11425} (\bibinfo {year} {2024})}\BibitemShut
  {NoStop}%
\bibitem [{\citenamefont {Abdurazakov}\ \emph {et~al.}(2023)\citenamefont
  {Abdurazakov}, \citenamefont {Li},\ and\ \citenamefont
  {Shim}}]{PhysRevB.108.125435}%
  \BibitemOpen
  \bibfield  {author} {\bibinfo {author} {\bibfnamefont {O.}~\bibnamefont
  {Abdurazakov}}, \bibinfo {author} {\bibfnamefont {C.}~\bibnamefont {Li}},\
  and\ \bibinfo {author} {\bibfnamefont {Y.-P.}\ \bibnamefont {Shim}},\
  }\bibfield  {title} {\bibinfo {title} {Formation of dark excitons in
  monolayer transition metal dichalcogenides by a vortex beam: Optical
  selection rules},\ }\href {https://doi.org/10.1103/PhysRevB.108.125435}
  {\bibfield  {journal} {\bibinfo  {journal} {Phys. Rev. B}\ }\textbf {\bibinfo
  {volume} {108}},\ \bibinfo {pages} {125435} (\bibinfo {year}
  {2023})}\BibitemShut {NoStop}%
\bibitem [{\citenamefont {Burkard}\ \emph {et~al.}(2023)\citenamefont
  {Burkard}, \citenamefont {Ladd}, \citenamefont {Pan}, \citenamefont
  {Nichol},\ and\ \citenamefont {Petta}}]{RevModPhys.95.025003}%
  \BibitemOpen
  \bibfield  {author} {\bibinfo {author} {\bibfnamefont {G.}~\bibnamefont
  {Burkard}}, \bibinfo {author} {\bibfnamefont {T.~D.}\ \bibnamefont {Ladd}},
  \bibinfo {author} {\bibfnamefont {A.}~\bibnamefont {Pan}}, \bibinfo {author}
  {\bibfnamefont {J.~M.}\ \bibnamefont {Nichol}},\ and\ \bibinfo {author}
  {\bibfnamefont {J.~R.}\ \bibnamefont {Petta}},\ }\bibfield  {title} {\bibinfo
  {title} {Semiconductor spin qubits},\ }\href
  {https://doi.org/10.1103/RevModPhys.95.025003} {\bibfield  {journal}
  {\bibinfo  {journal} {Rev. Mod. Phys.}\ }\textbf {\bibinfo {volume} {95}},\
  \bibinfo {pages} {025003} (\bibinfo {year} {2023})}\BibitemShut {NoStop}%
\bibitem [{\citenamefont {Zhu}\ \emph {et~al.}(2014)\citenamefont {Zhu},
  \citenamefont {Zhang}, \citenamefont {Glazov}, \citenamefont {Urbaszek},
  \citenamefont {Amand}, \citenamefont {Ji}, \citenamefont {Liu},\ and\
  \citenamefont {Marie}}]{PhysRevB.90.161302}%
  \BibitemOpen
  \bibfield  {author} {\bibinfo {author} {\bibfnamefont {C.~R.}\ \bibnamefont
  {Zhu}}, \bibinfo {author} {\bibfnamefont {K.}~\bibnamefont {Zhang}}, \bibinfo
  {author} {\bibfnamefont {M.}~\bibnamefont {Glazov}}, \bibinfo {author}
  {\bibfnamefont {B.}~\bibnamefont {Urbaszek}}, \bibinfo {author}
  {\bibfnamefont {T.}~\bibnamefont {Amand}}, \bibinfo {author} {\bibfnamefont
  {Z.~W.}\ \bibnamefont {Ji}}, \bibinfo {author} {\bibfnamefont {B.~L.}\
  \bibnamefont {Liu}},\ and\ \bibinfo {author} {\bibfnamefont {X.}~\bibnamefont
  {Marie}},\ }\bibfield  {title} {\bibinfo {title} {Exciton valley dynamics
  probed by kerr rotation in ${\mathrm{wse}}_{2}$ monolayers},\ }\href
  {https://doi.org/10.1103/PhysRevB.90.161302} {\bibfield  {journal} {\bibinfo
  {journal} {Phys. Rev. B}\ }\textbf {\bibinfo {volume} {90}},\ \bibinfo
  {pages} {161302} (\bibinfo {year} {2014})}\BibitemShut {NoStop}%
\bibitem [{\citenamefont {Song}\ \emph {et~al.}(2016)\citenamefont {Song},
  \citenamefont {Xie}, \citenamefont {Kang}, \citenamefont {Park},\ and\
  \citenamefont {Sih}}]{nanolett01727}%
  \BibitemOpen
  \bibfield  {author} {\bibinfo {author} {\bibfnamefont {X.}~\bibnamefont
  {Song}}, \bibinfo {author} {\bibfnamefont {S.}~\bibnamefont {Xie}}, \bibinfo
  {author} {\bibfnamefont {K.}~\bibnamefont {Kang}}, \bibinfo {author}
  {\bibfnamefont {J.}~\bibnamefont {Park}},\ and\ \bibinfo {author}
  {\bibfnamefont {V.}~\bibnamefont {Sih}},\ }\bibfield  {title} {\bibinfo
  {title} {Long-lived hole spin/valley polarization probed by kerr rotation in
  monolayer wse2},\ }\href {https://doi.org/10.1021/acs.nanolett.6b01727}
  {\bibfield  {journal} {\bibinfo  {journal} {Nano Lett.}\ }\textbf {\bibinfo
  {volume} {16}},\ \bibinfo {pages} {5010} (\bibinfo {year}
  {2016})}\BibitemShut {NoStop}%
\bibitem [{\citenamefont {Liu}\ \emph {et~al.}(2019)\citenamefont {Liu},
  \citenamefont {van Baren}, \citenamefont {Taniguchi}, \citenamefont
  {Watanabe}, \citenamefont {Chang},\ and\ \citenamefont
  {Lui}}]{PhysRevResearch.1.032007}%
  \BibitemOpen
  \bibfield  {author} {\bibinfo {author} {\bibfnamefont {E.}~\bibnamefont
  {Liu}}, \bibinfo {author} {\bibfnamefont {J.}~\bibnamefont {van Baren}},
  \bibinfo {author} {\bibfnamefont {T.}~\bibnamefont {Taniguchi}}, \bibinfo
  {author} {\bibfnamefont {K.}~\bibnamefont {Watanabe}}, \bibinfo {author}
  {\bibfnamefont {Y.-C.}\ \bibnamefont {Chang}},\ and\ \bibinfo {author}
  {\bibfnamefont {C.~H.}\ \bibnamefont {Lui}},\ }\bibfield  {title} {\bibinfo
  {title} {Valley-selective chiral phonon replicas of dark excitons and trions
  in monolayer $\mathrm{WS}{\mathrm{e}}_{2}$},\ }\href
  {https://doi.org/10.1103/PhysRevResearch.1.032007} {\bibfield  {journal}
  {\bibinfo  {journal} {Phys. Rev. Res.}\ }\textbf {\bibinfo {volume} {1}},\
  \bibinfo {pages} {032007} (\bibinfo {year} {2019})}\BibitemShut {NoStop}%
\bibitem [{\citenamefont {Yu}\ and\ \citenamefont
  {Wu}(2014)}]{PhysRevB.89.205303}%
  \BibitemOpen
  \bibfield  {author} {\bibinfo {author} {\bibfnamefont {T.}~\bibnamefont
  {Yu}}\ and\ \bibinfo {author} {\bibfnamefont {M.~W.}\ \bibnamefont {Wu}},\
  }\bibfield  {title} {\bibinfo {title} {Valley depolarization due to
  intervalley and intravalley electron-hole exchange interactions in monolayer
  {MoS}$_{2}$},\ }\href {https://doi.org/10.1103/PhysRevB.89.205303} {\bibfield
   {journal} {\bibinfo  {journal} {Phys. Rev. B}\ }\textbf {\bibinfo {volume}
  {89}},\ \bibinfo {pages} {205303} (\bibinfo {year} {2014})}\BibitemShut
  {NoStop}%
\bibitem [{\citenamefont {{D'yakonov}}\ \emph {et~al.}(1986)\citenamefont
  {{D'yakonov}}, \citenamefont {{Marushchak}}, \citenamefont {{Perel'}},\ and\
  \citenamefont {{Titkov}}}]{1986JETP63655D}%
  \BibitemOpen
  \bibfield  {author} {\bibinfo {author} {\bibfnamefont {M.~I.}\ \bibnamefont
  {{D'yakonov}}}, \bibinfo {author} {\bibfnamefont {V.~A.}\ \bibnamefont
  {{Marushchak}}}, \bibinfo {author} {\bibfnamefont {V.~I.}\ \bibnamefont
  {{Perel'}}},\ and\ \bibinfo {author} {\bibfnamefont {A.~N.}\ \bibnamefont
  {{Titkov}}},\ }\bibfield  {title} {\bibinfo {title} {{The effect of strain on
  the spin relaxation of conduction electrons in Ill-V semiconductors}},\
  }\href {https://doi.org/abs/1986JETP...63..655D} {\bibfield  {journal}
  {\bibinfo  {journal} {Sov. Phys. JETP}\ }\textbf {\bibinfo {volume} {63}},\
  \bibinfo {pages} {655} (\bibinfo {year} {1986})}\BibitemShut {NoStop}%
\end{thebibliography}
%

\end{document}